\documentclass[a4paper,12pt]{article}
\usepackage[utf8]{inputenc}
\pdfoutput=1 

\usepackage{jheppub} 

\usepackage[T1]{fontenc} 

\usepackage{graphicx,hyperref,color} 
\usepackage{amsmath,amssymb}
\usepackage{color}
\usepackage[italicdiff]{physics}
\usepackage{bm}
\usepackage{mathtools}
\usepackage{enumitem}
\usepackage{tensor}
\usepackage{wrapfig}
\usepackage{footnote}
\usepackage{here, comment}
\usepackage{shorttoc}
\usepackage{bbold}
\usepackage[normalem]{ulem}
\usepackage[export]{adjustbox}
\usepackage{bbm}
\usepackage[colorinlistoftodos]{todonotes}
\usepackage{comment}

\title{\boldmath 
Exploring causality in braneworld/cutoff holography via holographic scattering
}

\author[a,b]{Takato Mori}
\author[a]{Beni Yoshida}

\affiliation[a]{Perimeter Institute for Theoretical Physics, Waterloo, Ontario N2L 2Y5, Canada}
\affiliation[b]{
Center for Gravitational Physics and Quantum Information,
Yukawa Institute for Theoretical Physics, Kyoto University,\\ 
Kitashirakawa Oiwakecho, Sakyo-ku, Kyoto 606-8502, Japan}

\emailAdd{takato.mori@yukawa.kyoto-u.ac.jp}
\emailAdd{byoshida@perimeterinstitute.ca}

\abstract{
Holography with branes and/or cutoff surfaces presents a promising approach to studying quantum gravity beyond asymptotically anti-de Sitter spacetimes. However, this generalized holography is known to face several inconsistencies, including potential violations of causality and fundamental entropic inequalities. In this work, we address these challenges by investigating the bulk scattering process and its holographic realization. Specifically, we propose that the information on a brane/cutoff surface $Q$ propagates according to the induced light cones originating from a fictitious asymptotic boundary behind $Q$, rather than the conventional ones originating from a point on $Q$. Additionally, we establish the validity of the connected wedge theorem for generalized holography with induced light cones. We also demonstrate that entropic inequalities remain valid within the induced causal diamonds. While the induced light cone seemingly permits superluminal signaling, we argue that this causality violation can be an artifact of state preparation for radially propagating excitations, rather than local operator excitations on $Q$. 
}

\begin{document} 

\begin{flushright}
YITP-23-94
\end{flushright}

\maketitle

\flushbottom

\section{Introduction}\label{sec:intro}

In recent years, significant progress has been made in our understanding of quantum gravity within the realm of the AdS/CFT correspondence~\cite{Maldacena:1997re}. Nevertheless, an ultimate goal would be to promote the insights from the AdS/CFT correspondence and resolve mysteries in our own universe. In doing so, we would like to go beyond the anti-de Sitter (AdS) spacetime without spoiling its good controllability~\cite{Bousso:2022hlz,Bousso:2023sya,Nomura:2016ikr}.

A promising approach, called \emph{braneworld holography}, considers a universe on a brane $Q$ floating in the asymptotically AdS bulk. By generalizing holography to encompass a gravitating system~\cite{Karch:2000ct,Randall:1999ee,Randall:1999vf,Iwashita:2006zj}, this framework allows for the coexistence of both a conformal field theory (CFT) and a gravitating brane $Q$.  
Furthermore, it has been unveiled that a holographic boundary CFT (BCFT) is dual to the bulk spacetime terminating on the brane~\cite{Karch:2000gx,Aharony:2003qf}. 
Within this framework, a profound connection emerges among three entities: the bulk AdS spacetime denoted as $\mathcal{M}$, the holographic BCFT residing on $\Sigma$, and an intermediate picture where the braneworld $Q$ is coupled to a non-gravitating bath CFT located on the asymptotic boundary $\Sigma$~\cite{Takayanagi:2011zk,Fujita:2011fp,Almheiri:2019hni,Suzuki:2022xwv,Neuenfeld:2021wbl,Neuenfeld:2021bsb,Omiya:2021olc,Karch:2022rvr}.\footnote{We assume $\partial\mathcal{M}=Q\cup\Sigma$, and the neglected edge term plays no role in this paper.} This triangle equivalence is often called double holography and has also played a key role in recent studies of  evaporating black holes~\cite{Almheiri:2019psf,Almheiri:2019qdq,Almheiri:2019hni,Geng:2020qvw,Geng:2020fxl}.

Another promising approach, closely related to braneworld holography, is \emph{cutoff holography}. This approach begins with the conventional holographic setup featuring an AdS boundary, and then introduces a finite radial cutoff to remove the boundary. In recent years, cutoff AdS has garnered significant attention~\cite{McGough:2016lol, Kraus:2018xrn, Guica:2019nzm, Taylor:2018xcy, Hartman:2018tkw, Grado-White:2020wlb,Park:2018snf,Ota:2019yfe,Murdia:2019fax}. 
Notably, there is a strong belief that radially pushing the cutoff surface is dual to the $T\overline{T}$ deformation~\cite{Zamolodchikov:2004ce, Cavaglia:2016oda, Jiang:2019epa, Bonelli:2018kik,Donnelly:2018bef,Chen:2018eqk,Jeong:2019ylz,He:2019vzf,Banerjee:2019ewu,Chen:2019mis,Jafari:2019qns}. Cutoff holography finds additional motivation from the surface/state correspondence~\cite{Miyaji:2015yva} in tensor network toy models, where a wavefunction obeying the Hubeny-Rangamani-Takayanagi (HRT) formula~\cite{Ryu:2006bv,Ryu:2006ef,Hubeny:2007xt} can be constructed on any cutoff surface $\mathcal{Q}$. The rich interplay between tensor networks and holography has spurred further investigations into this approach, seeking a deeper understanding of its implications and potential applications.

Despite the elegance of these approaches, it has become evident that generalized holography faces certain inconsistencies including potential violations of causality and fundamental laws of quantum mechanics. These challenges include issues such as the violation of subadditivity/strong subadditivity~\cite{Lewkowycz:2019xse, Grado-White:2020wlb} as well as the occurrence of superluminal signaling due to bulk shortcuts~\cite{Omiya:2021olc,Greene:2011fm,Greene:2022urm,Greene:2022uyf,Polychronakos:2022uam,Dai:2023zsx}. Furthermore, this paper will delve into another puzzling aspect concerning brane causality and entanglement by exploring holographic quantum tasks in the bulk.
Elucidating and resolving these puzzles, as well as understanding the potential limitations of generalized holography, represent crucial initial steps towards drawing valuable lessons from the AdS/CFT correspondence and addressing fundamental questions beyond the correspondence~\cite{Geng:2023iqd}.

The underlying motivation behind this paper is two-fold. In the first part, we address the issue, concerning causality and entanglement in a setup with a brane/cutoff surface, by a certain versatile quantum information theory technique.
Specifically, we will utilize the \emph{holographic quantum task} paradigm, originally introduced by May~\cite{May:2019yxi}, in order to sharply identify and formulate the puzzle.
The key idea involves investigating information processing tasks through a bulk scattering process and finding the corresponding dual CFT realization. 
This approach allows us to demonstrate the existence of quantum entanglement in certain regions of the AdS boundary.
This result, known as the \emph{connected wedge theorem}, has been proven through both quantum information theoretic and gravitational techniques~\cite{May:2019odp}. 

Using this framework, we establish that the bulk scattering process involving a brane appears to be impossible due to the apparent breakdown of the connected wedge theorem, which would otherwise guarantee the presence of necessary entanglement.
This new puzzle highlights an apparent inconsistency in generalized holography that emerges from consideration of spacetime causality and quantum entanglement.

The second part of the paper presents resolutions to the aforementioned puzzle in holographic scattering, as well as other puzzles in generalized holography. Specifically, we address the following three puzzles: 
\begin{itemize}
\item Apparent breakdown of the connected wedge theorem in a holographic scattering process. 
\item Apparent violation of subadditivity (and strong subadditivity) for subregions involving a brane/cutoff surface. 
\item An apparent superluminal signaling in braneworld holography due to a bulk shortcut.
\end{itemize}

We propose resolutions to all three puzzles on an equal footing, based on the observation that the bulk local excitation on a brane/cutoff surface requires some initial preparation. 
To elaborate, we consider a fictitious asymptotic AdS boundary behind the brane/cutoff surface and examine the induced light cones on them. This naturally leads to an apparent superluminal causal diamond, which is seemingly acausal.
However, we emphasize that this does not violate causality, as the apparent superluminal propagation occurs due to careful initial preparation.
 
Specifically, by considering induced light cones from the fictitious boundary, we obtain the following resolutions. 

\begin{itemize}
\item The connected wedge theorem holds for induced light cones. This will be proven on a general ground from a geometric argument.
\item Subadditivity and strong subadditivity of holographic entanglement entropy~\cite{Ryu:2006bv,Ryu:2006ef,Hubeny:2007xt} remain valid when we use the induced causal diamond. This will also be proven on a general ground from a geometric argument.
\item An apparent superluminal signaling due to the bulk shortcut results from the induced light cones.
\end{itemize}

Our proposal naturally distinguishes two types of local excitations: \emph{trapped excitations} confined to the brane or cutoff surface and \emph{radially propagating excitations} which travels into the bulk. 
While trapped excitations follow standard causality and are created by local operators, radially propagating excitations require nonlocal operators and seemingly violate causality. 
This apparent causality violation is consistent with the HRT formula but only emerges when considering both types of excitations. 
We argue that the violation of subadditivity arises from conflicting assumptions of standard causality and the HRT formula. 

Holographic theory on branes and/or cutoff surfaces may become nonlocal due to higher derivative terms in its effective action after integrating out the bulk. 
Furthermore, the higher curvature corrections in the induced gravity and a finite effective UV cutoff on the brane indicate a breakdown of the semiclassical Einstein gravity approximation on the brane~\cite{Chen:2020uac,Chen:2020hmv}. 
These observations are often associated with a breakdown of the local Hilbert space structure on the brane/cutoff surface. 

Nevertheless, there is evidence to support the consideration of holographic theory with a local Hilbert space. 
In tensor network models, an arbitrary cut by a surface defines a state on that cut. 
The resulting wavefunction generally satisfies the Ryu-Takayanagi (RT) formula, suggesting that there is no fundamental obstacle in writing a wavefunction with a local Hilbert space on the cut. 
The results presented in this paper, particularly those concerning the connected wedge theorem and the (avoidance of) violations of entropic inequalities, further suggest that subregions defined by induced light cones possess well-defined operational meanings. 
After all, the effective UV cutoff in this context corresponds to a length scale below which we poorly understand the holographic theory and is not a fundamental obstacle in considering more fine-grained quantum mechanical degrees of freedom.  
In principle, nothing would prevent us from considering a local Hilbert space structure up to the Planck scale.
Consequently, we will adopt the perspective that while a local Hilbert space structure can still be defined, causal phenomena may propagate superluminally.

It is worth making a brief comment on different choices of boundary conditions. 
The choice between the Neumann boundary condition in braneworld holography and the Dirichlet boundary condition in cutoff holography results in distinct consequences.
The Neumann boundary condition on a brane $Q$ determines its profile, incorporating both the graviton and matter degrees of freedom. In contrast, imposing the Dirichlet boundary condition on a radial cutoff surface $\mathcal{Q}$ results in vanishing graviton fluctuations, leading to a purely quantum mechanical system on $\mathcal{Q}$ on a fixed curved background. Throughout most of the paper, we will assume the absence of graviton fluctuations by fixing the location of $Q$ or $\mathcal{Q}$. Therefore, we will interchangeably use the terms 'brane $Q$ and 'cutoff surface $\mathcal{Q}$. We will revisit this issue in Section~\ref{sec:bdy-cond}, where we discuss backreactions on $Q$ and its connection to the apparent breakdown of locality.

This paper is organized as follows.
\begin{itemize}
\item In Section~\ref{sec:AdS-no_brane}, we present a brief review of a conceptual puzzle concerning the bulk scattering process in the AdS/CFT and its resolution by the holographic quantum task paradigm. We also review related developments on the connected wedge theorem. 

\item In Section~\ref{sec:hol-task-brane}, we examine the bulk scattering process in the presence of a brane $Q$ and demonstrate that the connected wedge theorem does not hold in this setup. We also demonstrate that holographic entanglement entropy on a brane $Q$ violates subadditivity of entanglement. 

\item In Section~\ref{sec:induced}, we introduce the notion of induced light cones and present an overview of our proposals for resolving the aforementioned puzzles related to causality and entanglement in setups with a brane $Q$ and cutoff surface $\mathcal{Q}$. 

\item In Section~\ref{sec:CWT}, we generalize the connected wedge theorem to setups with $Q$ or $\mathcal{Q}$ by considering induced light cones which appear to propagate superluminally. 

\item In Section~\ref{sec:SA-ind}, we show that holographic entanglement entropy obeys basic quantum information inequalities, such as subadditivity and strong subadditivity, by considering induced causal diamonds which appear to spread superluminally. 

\item In Section~\ref{sec:OW-puzzle}, we briefly comment on the puzzle by Omiya and Wei concerning superluminal signaling in setups with $Q$ and $\mathcal{Q}$. 

\item In Section~\ref{sec:proposals}, we summarize our proposal and its physical interpretation. We also sketch tensor network realizations of wavefunctions with $Q$ or $\mathcal{Q}$. We conclude by revisiting the boundary condition problem. 
\end{itemize}

\noindent\textbf{Note added}: 
During the preparation of this manuscript, we became aware of an independent work by Dominik Neuenfeld and Manu Srivastava which pointed out a similar setup for violation of subadditivity on a brane~\cite{Neuenfeld:2023svs}. 

\section{Review of connected wedge theorem}\label{sec:AdS-no_brane} 

The AdS/CFT correspondence asserts that the bulk quantum gravity has a holographic realization on the boundary quantum system. How this can be possible, however, often poses important conceptual puzzles. Here we present a brief review of recent important findings concerning scattering events in the bulk and their implications on the entanglement structure of the dual CFT~\cite{May:2019yxi}.
In this paper, we focus on the cases with the three-dimensional bulk and the two-dimensional quantum field theory (QFT) dual.

To illustrate the idea, consider information processing between Alice and Bob via direct scattering of signals in the bulk.
The pure global AdS${}_3$ metric is given by
\begin{equation}
    ds^2 = -(R^2+r^2)dt^2 +\frac{R^2}{R^2+r^2}dr^2 + r^2 d\theta^2,
    \label{eq:globalAdS}
\end{equation}
where $R$ is the AdS radius. The dual CFT lives on the asymptotic boundary $\Sigma$ at $r=r_\infty \gg R$, where $\epsilon=R^2/r_\infty$ is the UV cutoff. It will be useful to write it in terms of the embedding coordinates $(X_0,X_3,X_1,X_2)$ and Poincar\'e coordinates $(\tau,z,x)$ :
\begin{equation}
    \begin{alignedat}{3}
        X_0 &= &\sqrt{R^2+r^2}\cos t\ &= &\frac{R^2+z^2+x^2-\tau^2}{2z},& \\
        X_3 &= &\sqrt{R^2+r^2}\sin t\ &= &\frac{R\tau}{z},\phantom{\frac{x-\tau^2}{2z}}& \\
        X_1 &= &r\sin\theta &= &\frac{Rx}{z},\phantom{\frac{x-\tau^2}{2z}}& \\
        X_2 &= &-r\cos\theta &= &\frac{-R^2+z^2+x^2-\tau^2}{2z}&.
        \label{eq:poincare-global}
    \end{alignedat}
\end{equation}
Then, the metric is given by
\begin{equation}
    ds^2= -dX_0^2-dX_3^2 +dX_1^2 +dX_2^2.
\end{equation}

\begin{figure}
    \centering
    \includegraphics[width=\linewidth]{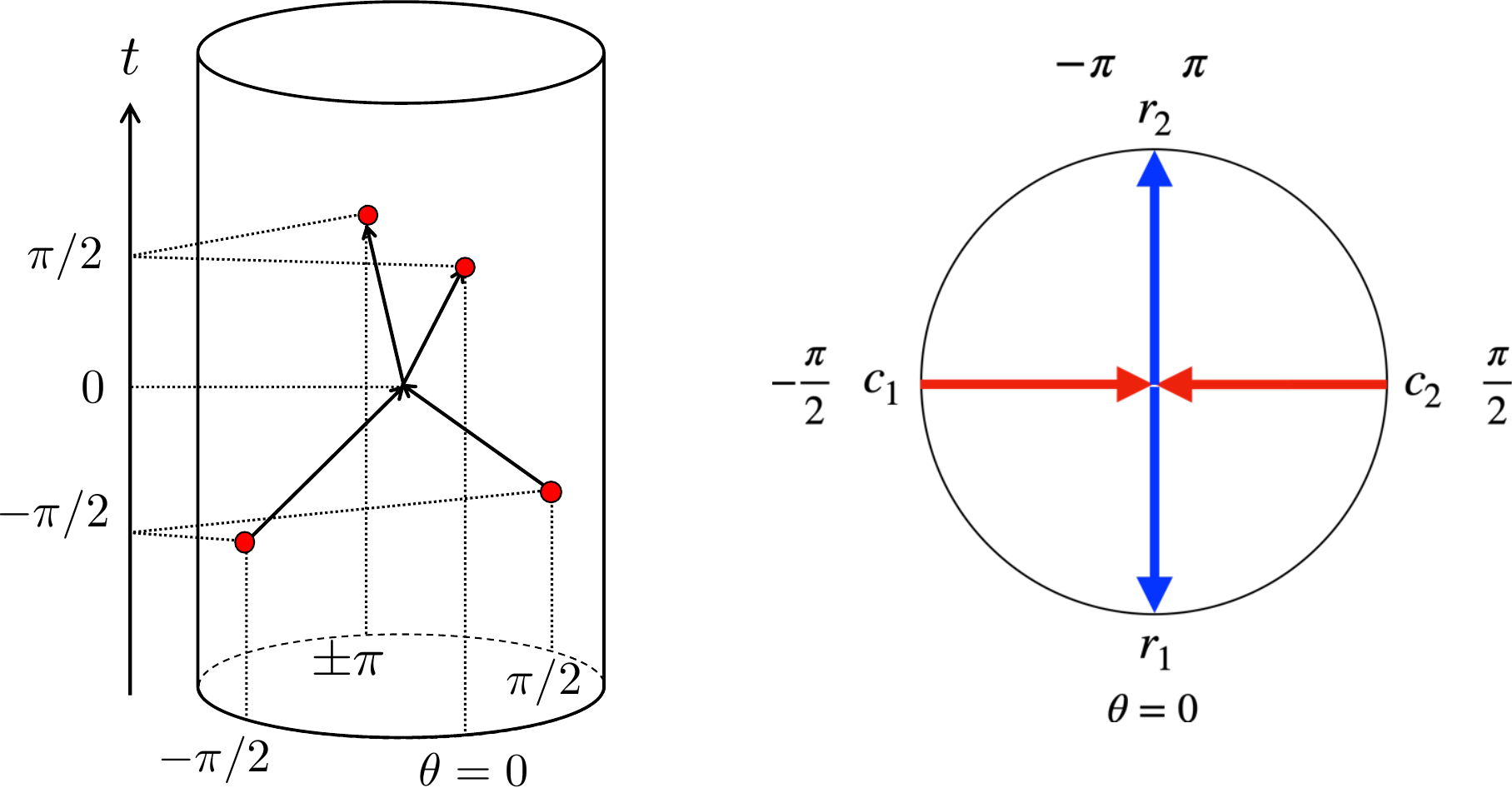}
    \caption{An example of 2-to-2 null scattering events whose input points are $c_1:(t=-\pi/2,\theta=-\pi/2)$ and $c_2:(-\pi/2,\pi/2)$ and output points are $r_1:(\pi/2,0)$ and $r_2:(\pi/2,\pi)$. The right figure shows the top view.}
    \label{fig:scat-wo-brane-diag}
\end{figure}

Let us focus on an asymptotic scattering process where the input points $c_1,c_2$ are located at $\theta=-\frac{\pi}{2}, +\frac{\pi}{2}$ while the output points $r_1,r_2$ are located at $\theta = 0,\pi$ as shown in Fig.~\ref{fig:scat-wo-brane-diag}. 
When the input and output points are separated by $\Delta t = \pi$ in the boundary time, two signals from $c_1,c_2$ can meet at the center of the AdS bulk and then scatter off to $r_1,r_2$.
Formally, the bulk scattering region can be identified by considering intersections of bulk domains of dependence, namely
\begin{align}
    P \equiv J^{+}\left(c_{1}\right) \cap J^{+}\left(c_{2}\right) \cap J^{-}\left(r_{1}\right) \cap J^{-}\left(r_{2}\right) \not=\emptyset,
\end{align}
where ${J}^{\pm}(p)$ represent the future/past \emph{bulk} domain of dependence of a point $p$.
In our setup, the scattering region $P$ consists of just a point at the center of the bulk.

\begin{figure}
    \centering
    \includegraphics[width=0.5\linewidth]{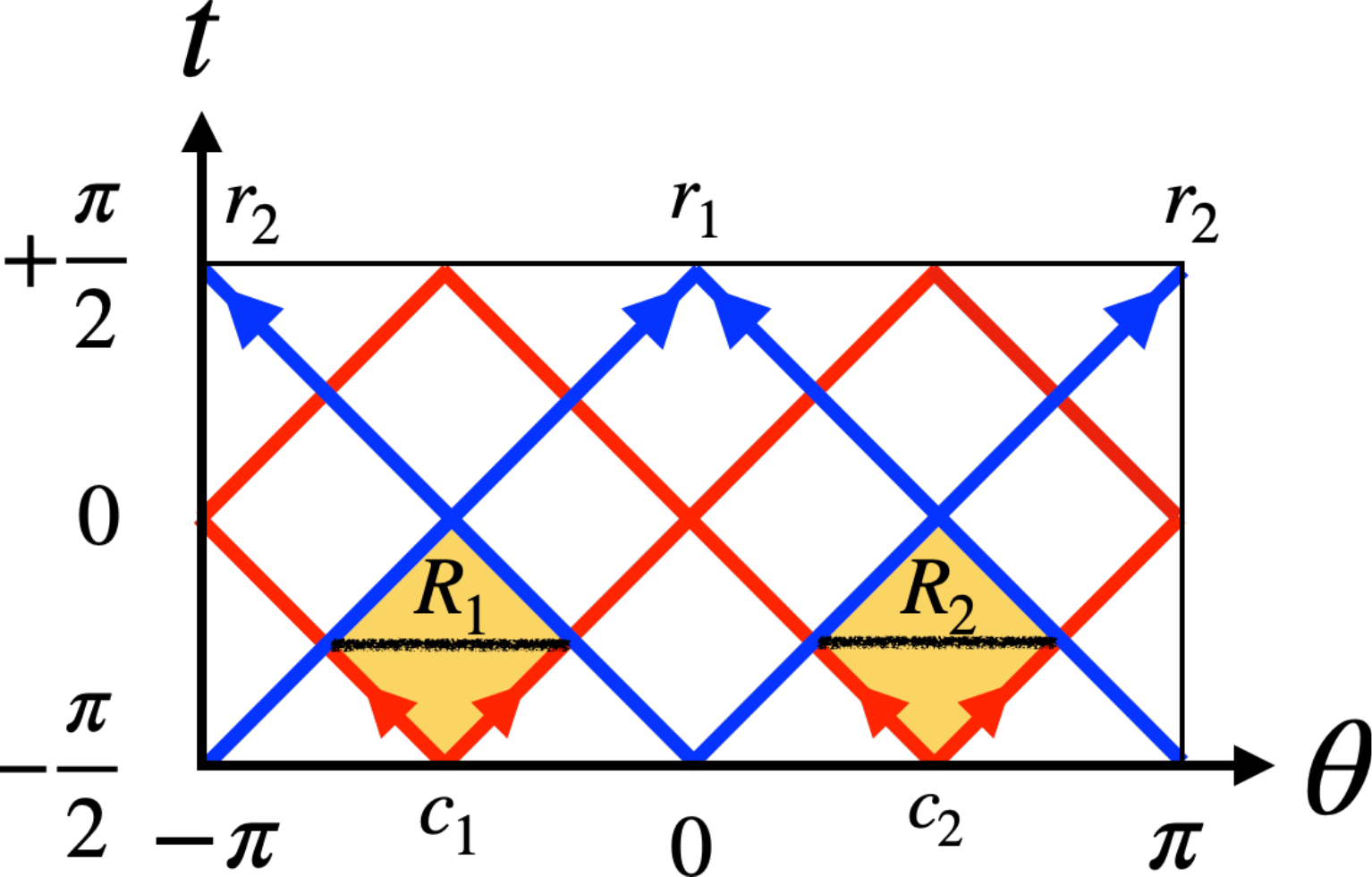}
    \caption{The light rays on the asymptotic boundary. . 
    The future light cones of $c_1,c_2$ are denoted by the red lines, and the past light cones of $r_1,r_2$ are denoted by the blue lines. The input regions $R_i$ are defined by the domains of dependence $D\left(R_{i}\right)=\hat{J}^{+}\left(c_{i}\right) \cap \hat{J}^{-}\left(r_{1}\right) \cap \hat{J}^{-}\left(r_{2}\right)$ colored in orange.}
    \label{fig:scat-wo-brane}
\end{figure}

Viewing the same scattering process from the boundary viewpoint, however, it is not possible to have direct scattering on the asymptotic boundary $\Sigma$ (Fig.~\ref{fig:scat-wo-brane}). 
The impossibility of direct boundary scattering can be formally confirmed by observing that the boundary scattering region is empty, namely
\begin{align}
    \hat{P} \equiv \hat{J}^{+}\left(c_{1}\right) \cap \hat{J}^{+}\left(c_{2}\right) \cap \hat{J}^{-}\left(r_{1}\right) \cap \hat{J}^{-}\left(r_{2}\right) = \emptyset,
\end{align}
where $\hat{J}^{\pm}(p)$ represent the future/past \emph{boundary} domain of dependence of a point $p$. 
Hence, it appears that direct scattering between two signals in the dual CFT cannot be holographically realized while the same task is allowed in the bulk. The corresponding causal structures are given in Fig.~\ref{fig:scat-wo-brane}. 

Fortunately, this puzzle can be avoided; indeed, it has been long known in quantum information theory that various kinds of two-party quantum information processing tasks can be realized even without direct contact by utilizing pre-shared quantum entanglement. 
A prime example is the \textbf{B}$_{84}$ task, which was originally considered in studies of quantum cryptography~\cite{Bennett_2014}. 
The goal of the \textbf{B}$_{84}$ task is to share some classical information $b$ between Alice and Bob when $b$ is initially prepared for Alice in some ``hidden'' form which can be revealed only by knowing Bob's initial state. 
Specifically, Alice's input is $H^q\ket{b}$ and Bob's input is $\ket{q}$, where $\ket{b}$ and $\ket{q}$ are independently- and randomly-chosen qubits in the $Z$-eigenbasis $\{\ket{0},\ket{1}\}$ and $H$ is the Hadamard gate. 
When Alice and Bob can meet directly (as in the bulk scattering process), Bob can simply tell the value of $q$ to Alice and reveal $b$ by applying $H^q$ to Alice's qubit. 
Here, the question is whether Alice and Bob can perform this task without a direct meeting (as in the boundary scattering process). 
The upshot is that the \textbf{B}$_{84}$ task can be indeed performed if Alice and Bob share an EPR pair by using quantum teleportation. 
Namely, Bob can nonlocally implement $H^q\ket{b}$ to Alice's qubit by teleporting Alice's qubit to Bob. 
By sending the Bell measurement outcome in quantum teleportation to output locations and applying appropriate feedback operations, the \textbf{B}$_{84}$ task can be performed without a direct meeting between Alice and Bob. 

In fact, one can promote this observation and provide a quantum information theoretic proof for the existence of quantum entanglement between boundary regions in the dual CFT. 
The key insight is that the success of the \textbf{B}$_{84}$ task necessitates a finite mutual information shared between Alice and Bob as shown in~\cite{Bennett_2014,May:2019yxi}. 
In particular, let us consider a family of tasks, denoted by \textbf{B}$_{84}^{\times n}$, which consists of the \textbf{B}$_{84}$ tasks repeated $n$ times in parallel. 
In order to execute \textbf{B}$_{84}^{\times n}$ with high success probability, Alice and Bob must possess a shared resource state $\rho_{AB}$ which satisfies 
\begin{equation}
    I(A:B)\equiv S_{A}+S_{B}-S_{A B}\ge 2\left(n \log \sec^2(\pi/8)-\log 2\right).
    \label{eq:MI-B84}
\end{equation}
One may choose $n$ large as long as $n<O(N^2)$ in order to avoid backreaction in the semiclassical treatment in the bulk. 
As such, the resource state must have at least $O(N^{a})$ mutual information with $a\approx 2$. 

What is the resource entanglement in the dual CFT that can be utilized in the scattering event? The regions of concern, called \emph{input regions}, can be identified from the intersections of the domains of dependence:
\begin{equation}
   D\left(R_{i}\right)\equiv \hat{J}^{+}\left(c_{i}\right) \cap \hat{J}^{-}\left(r_{1}\right) \cap \hat{J}^{-}\left(r_{2}\right),
    \label{eq:DoD-R}
\end{equation}
where $\hat{J}^\pm$ denotes the future/past domain of dependence on the asymptotic boundary. The input region $D(R_i)$
is accessible to one of the input points $c_i$, and is causally connected to both output points $r_1,r_2$ (see Fig.~\ref{fig:scat-wo-brane}). 
Recalling the HRT formula, saying that $R_1$ and $R_2$ can possess a large amount of mutual information when their entanglement wedge~\cite{Czech:2012bh,Headrick:2014cta,Wall:2012uf,Jafferis:2015del} is connected, we obtain the connected wedge theorem.\footnote{The boundary input regions $R_1$ and $R_2$ are non-overlapping since the boundary direct scattering is prohibited, namely $\hat{P}=\emptyset$.} 

\begin{itemize}[leftmargin=*]
\item[] \textbf{Connected wedge theorem:} Assume that the bulk scattering region is nonempty, but the boundary scattering region is empty:
\begin{align}
P \not= \emptyset, \qquad \hat{P}=\emptyset. \label{eq:condition}
\end{align}
Then, the input regions $R_1$ and $R_2$ must be entangled,\footnote{Precisely speaking, the amount of mutual information must be as large as $O(1/G_N^b)$, where $b\lesssim 1$.} and hence, their entanglement wedge must be connected.
\end{itemize}

For a particular arrangement of input/output points considered in Fig.~\ref{fig:scat-wo-brane-diag}, one can indeed verify that the entanglement wedge for $R_1$ and $R_2$ is connected (Fig.~\ref{fig:scat-wo-brane}). This powerful theorem states that the connected entanglement wedge is actually a generic feature whenever $c_1,c_2,r_1,r_2$ satisfy~\eqref{eq:condition}. The above argument, based on the \textbf{B}$_{84}$ task, provides a quantum information theoretic proof of the connected wedge theorem. It turns out that the same statement can be proven from a geometric argument as well~\cite{May:2019odp}. In Section~\ref{sec:geom-proof-CWthm}, we will review the geometric proof and further generalize the theorem to the setup with a brane and in cutoff holography.

\section{Holographic quantum task with a brane}\label{sec:hol-task-brane}

In this section, we examine whether the connected theorem holds in the intermediate picture in the case of pure global AdS${}_3$ with an end-of-the-world (ETW) brane $Q$. 
Namely, we demonstrate that the entanglement wedge for the input regions $R_1$ and $R_2$ is disconnected when one of the input signals emanates from the brane $Q$.
A similar argument holds for cutoff holography as well if one replaces the ETW brane $Q$ with an inhomogeneous cutoff surface $\mathcal{Q}$ with the same profile.

We would like to note that a previous work~\cite{May:2021zyu} studied a scattering process involving a brane in a different setup.
Specifically, they considered a situation where one of the inputs is distributed over the entire brane and studied a 1-to-2 scattering process. 
In this paper, we consider the 2-to-2 scattering process which allows us to study the entanglement wedge for subregions involving a brane $Q$ as opposed to the previous work.

\subsection{The AdS/BCFT correspondence}

We begin by reviewing the basic setup of the AdS/BCFT correspondence and introduce a few useful coordinate systems to describe a 2-to-2 scattering process between the brane $Q$ and the asymptotic boundary $\Sigma$. 

In the AdS/BCFT correspondence~\cite{Takayanagi:2011zk,Fujita:2011fp}, the bulk spacetime terminates on the end-of-the-world (ETW) brane $Q$, whose action is given by
\begin{equation}
    \frac{1}{8\pi G_N}\int_Q \sqrt{-h} (K-T),
\end{equation}
where $G_N$ is Newton's constant of the bulk gravity, $h$ is the determinant of the induced metric on $Q$, $K$ is the trace of the extrinsic curvature, and $T$ is a tension of the brane. 
Imposing the Neumann boundary condition on $Q$,
\begin{equation}
    K_{ab}-K h_{ab} = T h_{ab},
    \label{eq:neumann-bc}
\end{equation}
we find a simple solution 
\begin{equation}
    r\sin\theta=-\lambda R,
    \label{eq:brane-global}
\end{equation}
where the boundaries of the dual BCFT are located at $\theta=0,\pi$ and $\lambda$ is related to the tension $T$ as 
\begin{equation}
    TR=\frac{\lambda}{\sqrt{1+\lambda^2}}.
\end{equation}
Thus, $T=0$ corresponds to $\lambda=0$ and $T\rightarrow 1$ corresponds to $\lambda\rightarrow \infty$.\footnote{When $T\rightarrow 1$, the brane approaches the asymptotic boundary. In such a case, we need to carefully relate the UV cutoff $R^2/r_\infty$ to the limit. See Appendix~\ref{app:cutoff} for details.}

\begin{figure}
    \centering
    \includegraphics[width=0.5\linewidth,clip]{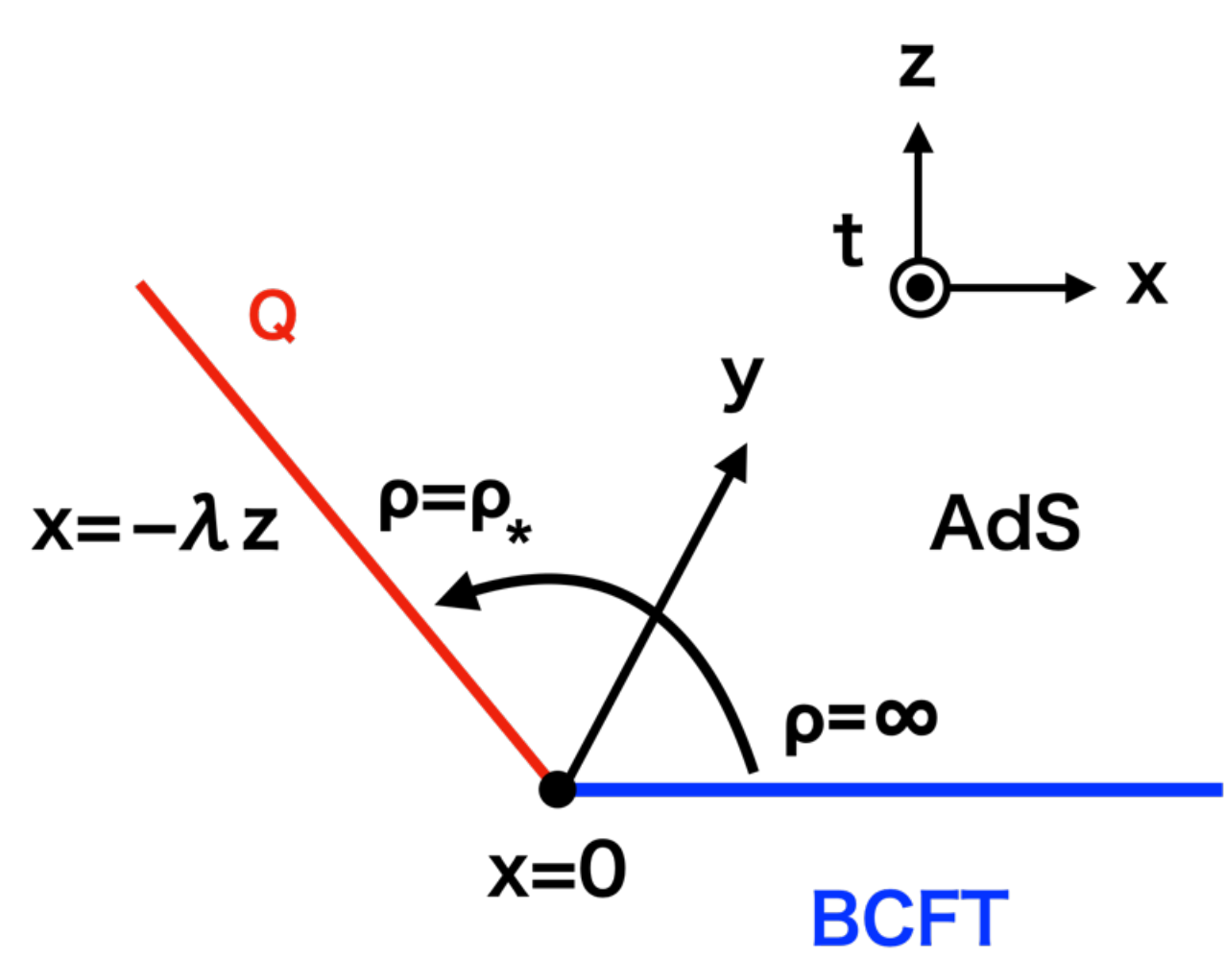}
    \caption{The gravity dual of a BCFT (blue line), whose boundary is located at $x=0$, is given by the Poincar\'e AdS bounded by the ETW brane (red line) $Q$. The brane is located at $\rho=\rho_\ast$ or $x=-\lambda z$.}
    \label{fig:poincare-AdS/BCFT}
\end{figure}

By a coordinate transformation from the Poincar\'e coordinates 
\begin{equation}
    x=y\tanh \frac{\rho}{R},\quad z= \frac{y}{\cosh\frac{\rho}{R}}, 
\end{equation}
where $y\ge 0$ and $\rho\in(-\infty,\infty)$, we obtain the following metric
\begin{equation}
    ds^2 = d\rho^2 + R^2 \cosh^2\frac{\rho}{R}\frac{-d\tau^2+dy^2}{y^2},
\end{equation}
where the brane profile is given by a constant-$\rho$ surface. The location of the brane $\rho=\rho_\ast$ is related to the tension $T$ by
\begin{equation}
    TR=-\tanh \frac{\rho_\ast}{R}.
\end{equation}
The setup in the Poincar\'e coordinates is shown in Fig.~\ref{fig:poincare-AdS/BCFT}.

\begin{figure}
    \centering
    \includegraphics[width=0.3\linewidth,clip]{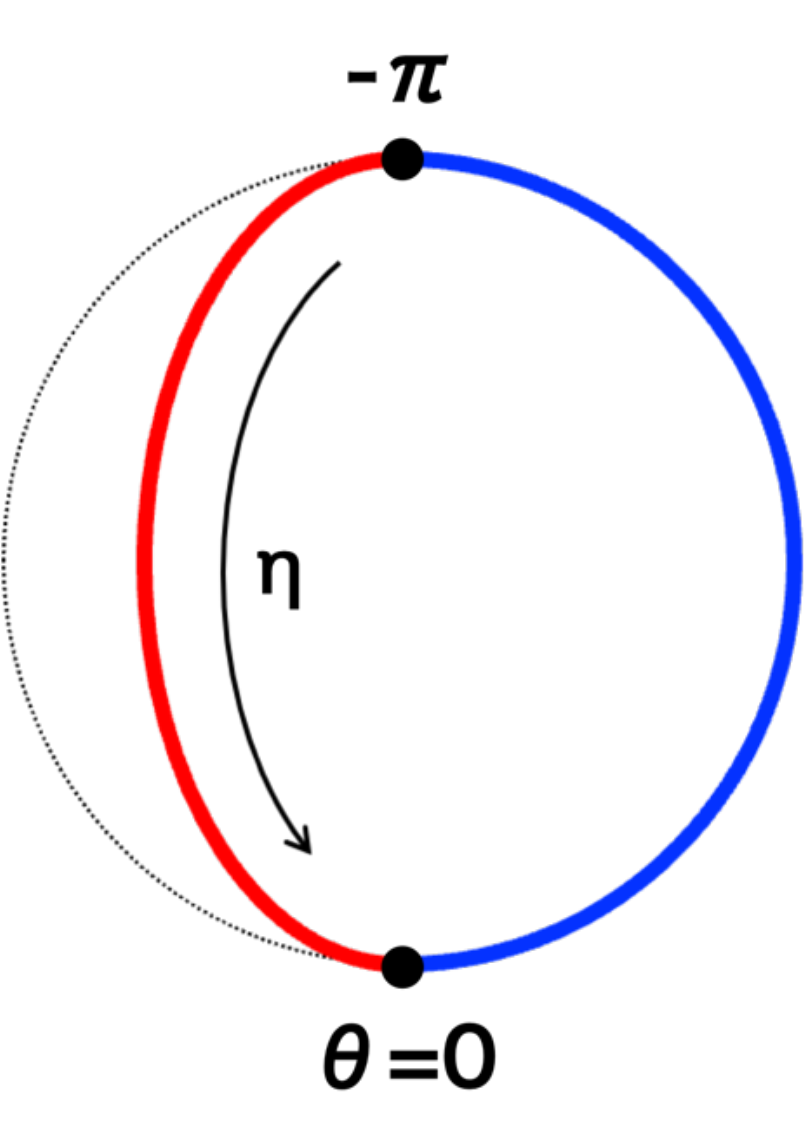}
    \caption{The coordinates on the brane for positive tension. $\abs{\eta}$ takes a value between $0$ and $\pi$ and its sign equals that of the slope $-\lambda$. The direction is counterclockwise, the same as $\theta$. }
    \label{fig:eta-coords}
\end{figure}

To discuss signals emanating from the brane $Q$, it is useful to introduce conformally-flat coordinates in global AdS such that
\begin{equation}
    ds^2\big|_Q= (r^2+R^2) (-dt^2+d\eta^2). \label{eq:eta-metric}
\end{equation}
The null trajectory along the brane is characterized by $t=\pm \eta$ plus constant.
It follows from \eqref{eq:eta-metric} that
\begin{equation}
    d\eta^2 = \frac{R^2}{(r^2+R^2)^2}dr^2 +\frac{r^2}{r^2+R^2}d\theta^2.
\end{equation}
We choose $\eta$ such that the sign of $\eta$ coincides with that of $\theta$ when the brane is close to the boundary ($T\rightarrow 1$) as depicted in Fig.~\ref{fig:eta-coords}:
\begin{equation}
    d\eta = d\theta \sqrt{\frac{R^2}{(r^2+R^2)^2}\qty(\frac{dr}{d\theta})^2 +\frac{r^2}{r^2+R^2}}.
\end{equation}

By plugging in the brane profile, $r\sin\theta=-\lambda R$ \eqref{eq:brane-global}, we obtain
\begin{align}
    d\eta &= d\theta \sqrt{\frac{\lambda^2}{\lambda^2+\sin^2\theta}\qty(1+\frac{\cos^2\theta}{\lambda^2+\sin^2\theta})} \nonumber\\
    \Rightarrow \eta & =  \arctan\qty(\sqrt{1+\frac{1}{\lambda^2}}\tan\theta), 
    \label{eq:eta-theta}
\end{align}
where we take $\arctan x\in (-\pi,0]$ for $T>0$ and $\arctan x\in [0,\pi)$ for $T<0$. 
Note that when $\eta\rightarrow 0,\pm\pi$, the limiting value of $\eta$ needs to be carefully related to the UV cutoff $R^2/r_\infty$ (Appendix~\ref{app:cutoff}). 
In the following, we focus on the $T>0$ case (\emph{i.e.} $\lambda>0$ and $\eta\le 0$).

In Poincar\'e coordinates, it follows from \eqref{eq:poincare-global} that
\begin{equation}
    \frac{y^2}{R^2}=z^2+x^2=\frac{R^2+r^2\sin^2\theta}{(\sqrt{R^2+r^2}\cos t + r\cos\theta)^2}.
\end{equation}
On the brane $Q$, it becomes
\begin{equation}
    \left.\frac{y^2}{R^2}\right|_Q = \frac{\sin^2\eta}{\qty(\cos t + \cos\eta)^2}.
    \label{eq:rel-y-eta}
\end{equation}
When $t=0$, it simplifies to
\begin{equation}
    \left.\frac{y^2}{R^2}\right|_Q = \frac{1-\cos^2\eta}{\qty(1 + \cos\eta)^2} = \frac{1-\cos\eta}{1+\cos\eta}=\tan^2\frac{\eta}{2}.
\end{equation}

\begin{figure}
    \centering
    \includegraphics[width=0.3\linewidth]{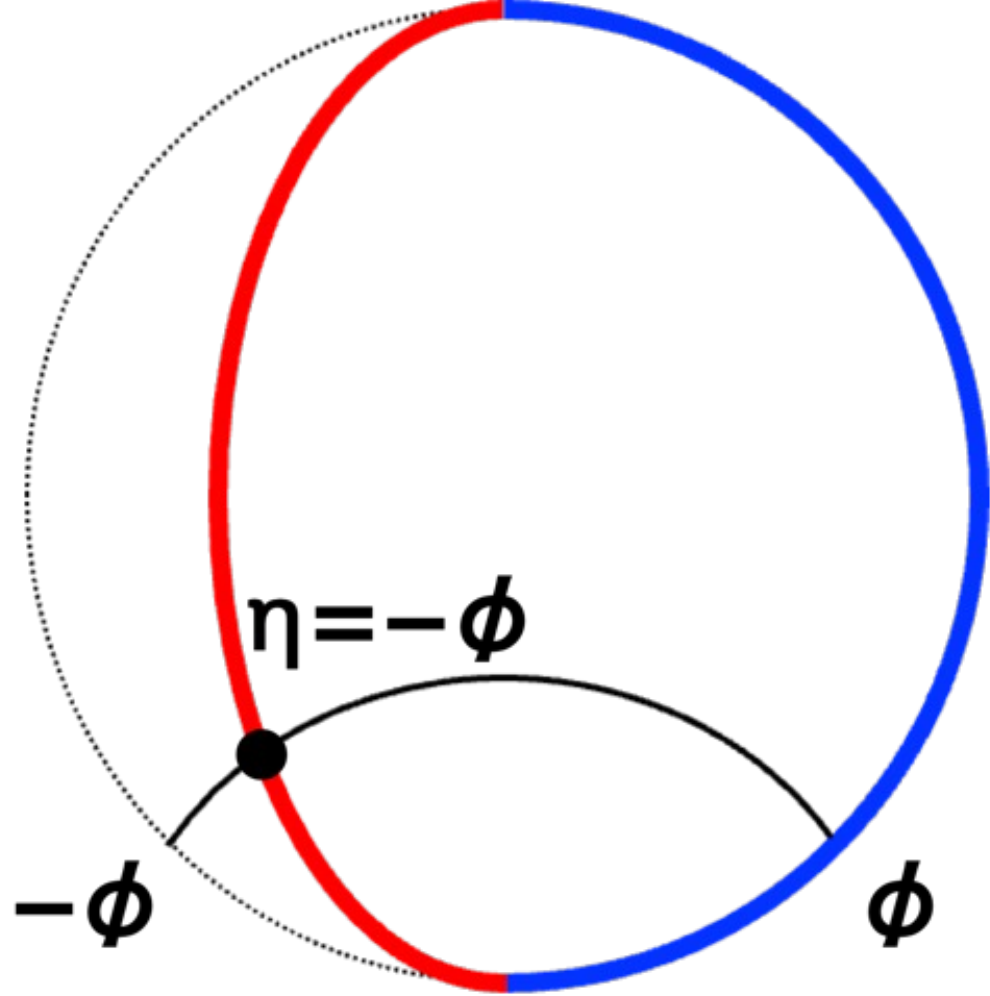}
    \caption{A geometric relation between the coordinates on the brane and the boundary. 
    When a geodesic intersects with the brane, $\theta$ at the endpoint of the geodesic equals $\eta$ at the intersection.
    }
    \label{fig:geodesic-eta}
\end{figure}

We conclude by presenting a useful geometric property concerning $\eta$. 
Consider a geodesic that is symmetric around $\theta=0$, and its intersection with $Q$, as depicted in Fig.~\ref{fig:geodesic-eta}.
When the geodesic extends from $x=-L$ to $L$ in the Poincare coordinates, the value of $y$ at the intersection of the brane and the geodesic is given by $y=L$. From \eqref{eq:rel-y-eta}, the corresponding value of $\eta=\eta(L)$ satisfies
\begin{equation}
    \frac{L^2}{R^2}=\frac{\sin^2\eta(L)}{\qty(\cos t +\cos \eta(L))^2}.
    \label{eq:eta-geo1}
\end{equation}
On the other hand, the $\theta$ coordinate $\theta(L)=\pm\phi$ at the asymptotic boundary corresponding to $x=\pm L$ satisfies
\begin{equation}
    \frac{\sin^2\theta(L)}{(\cos t +\cos \theta(L))^2}=\frac{(r\sin\theta(L))^2}{(r\cos t +r\cos \theta(L))^2}=\left.\frac{x^2}{R^2}\right|_{r\rightarrow \infty}=\frac{L^2}{R^2},
    \label{eq:eta-geo2}
\end{equation}
where we used $r\sim \sqrt{R^2+r^2}$ at the asymptotic boundary $r\rightarrow \infty$ in the second equality. Equating \eqref{eq:eta-geo1} and \eqref{eq:eta-geo2}, we obtain
\begin{equation}
    \eta(L)=-\phi=\theta(x=- L),
\end{equation}
where the sign is fixed from that of the tension.

\subsection{Holographic quantum task between brane and boundary}\label{sec:scat-brane-local}

\begin{figure}
\hspace*{-10pt}
    \centering
    \includegraphics[width=1.06\linewidth]{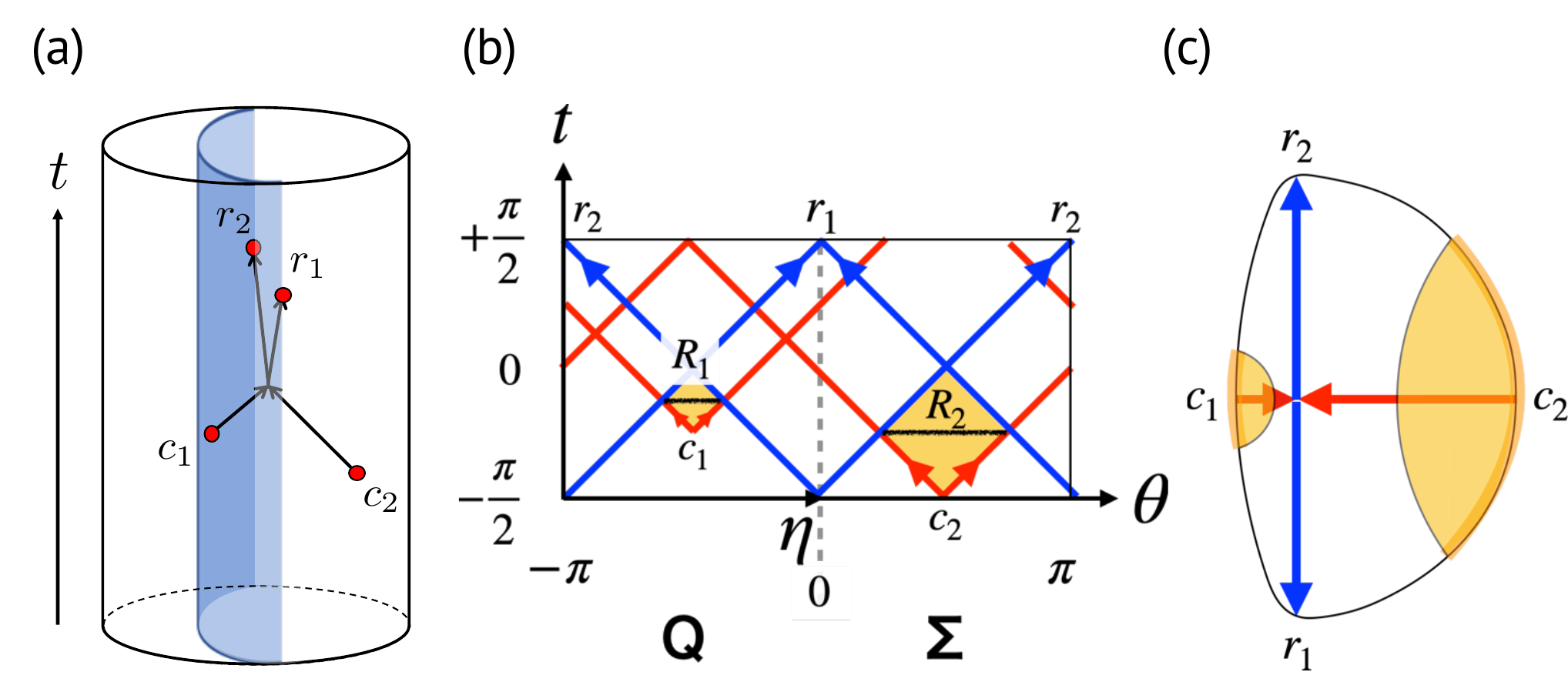}
    \caption{ 
    A 2-to-2 scattering event between the brane $Q$ and the asymptotic boundary $\Sigma$. (a) The bulk picture of the scattering. $t$ denotes the Lorentzian time. (b) The light cones for the brane input $c_1$, the boundary input $c_2$, the brane output $r_1$, and the boundary output $r_2$ in the intermediate picture $Q\cup\Sigma$. Each input region $R_{1,2}$ are defined by $D\left(R_{i}\right)=\hat{J}^{+}\left(c_{i}\right) \cap \hat{J}^{-}\left(r_{1}\right) \cap \hat{J}^{-}\left(r_{2}\right)$ colored in orange. (c) For any positive tensions, $R_1$ is too small to have a connected entanglement wedge between $R_1$ and $R_2$.}
    \label{fig:scat-brane-local}
\end{figure}

Let us consider the same scattering process as Section~\ref{sec:AdS-no_brane}, but with the ETW brane $Q$, where the scattering happens at the center of pure AdS (Fig.~\ref{fig:scat-brane-local} (a)). This requires $T>0$. We fix the input point $c_2$ and output point $r_2$ on the asymptotic boundary $\Sigma$. The output point $r_1$ is placed on the edge of the brane $\eta=0$.
The location of the input point $c_1$ on $Q$ is given by the intersection of
\begin{equation}
    \begin{cases}
    \theta&=-\pi/2\\
        r\sin\theta&=-\lambda R,\\
        -(r^2+R^2)dt^2+\displaystyle\frac{R^2}{r^2+R^2}dr^2&=0.
    \end{cases}
\end{equation}
Since we consider a signal emanating from $c_1$, the third equation reduces to $t=-\arctan(r/R)$. 
For $t$, we take the range of $\arctan$ to be $[-\pi/2,\pi/2]$. Then, 
$c_1$ is located at
\begin{equation}
    (t,\theta)=(-\arctan\lambda,-\pi/2)=(t,\eta).
\end{equation}

On the contrary, one can see that the boundary direct scattering is causally prohibited as shown in Fig.~\ref{fig:scat-brane-local} (b), and as such, two input regions $R_1,R_2$ must have access to shared quantum entanglement. 
Recall that input regions $R_1,R_2$ were defined from
\begin{equation}
    D(R_i)\equiv \hat{J}^{+}\left(c_{i}\right) \cap \hat{J}^{-}\left(r_{1}\right) \cap \hat{J}^{-}\left(r_{2}\right),
    \label{eq:DoD-R2}
\end{equation}
where $\hat{J}^\pm$ denotes the future/past domain of dependence on $Q\cup\Sigma$. Since the light cone on $Q$ is just given by $ds^2=0=-dt^2+d\eta^2$, 
the past light cones for $r_{1,2}$ are $t=\pm\eta\pm\pi/2$. $\partial R_1$ are their intersections with the future light cone of $c_1$, \emph{i.e.}
\begin{equation}
    (t,\eta)=\qty(-\frac{1}{2}\arctan\lambda, -\frac{\pi}{2}\pm \frac{1}{2}\arctan\lambda).
    \label{eq:partial-R1-usual}
\end{equation}
Similarly, $\partial R_2$ are
\begin{equation}
    (t,\theta)=\qty(-\frac{\pi}{4},\frac{\pi}{2}\pm\frac{\pi}{4}).
    \label{eq:partial-R2-usual}
\end{equation}
The light cones and the input regions $R_1$ and $R_2$ are illustrated in Fig.~\ref{fig:scat-brane-local} (b).

Whether a naive application of the HRT prescription for a subsystem on the brane is valid or not is a subtle issue, but let us assume that the holographic entanglement entropy can be indeed computed from the geodesics connecting two endpoints. 
Following the calculations given in Appendix \ref{app:HEE}, the length of the disconnected geodesics is calculated from \eqref{eq:zeta-discon-bdy} and \eqref{eq:zeta-discon-Q}, namely,
\begin{equation}
    \zeta^{dis}_{\Sigma}
    =\frac{r^2_\infty}{R^2}
    \label{eq:discon-geo-R2}
\end{equation}
and 
\begin{equation}
    \zeta^{dis}_Q 
    = 2(1+\lambda^2) \frac{\sqrt{1+\lambda^2}-1}{\sqrt{1+\lambda^2}+1}+1,
\end{equation}
where $\zeta$ is defined in \eqref{eq:zeta}. 
For connected geodesics, from \eqref{eq:zeta-con},
\begin{align}
    \zeta_{\Sigma Q}^{con}
    &=\frac{1}{\sqrt{2}}\frac{r_\infty}{R}\qty(\sqrt{1+\lambda^2} +\lambda).
\end{align}

The length of the disconnected geodesics is calculated from \eqref{eq:geodesic-gen} and \eqref{eq:geodesic-bdy-lim}:
\begin{align}
    \frac{d_{dis}}{R}
    &= 2\log \frac{r_\infty}{R} +\log 2 + 
    \log\qty(\zeta^{dis}_Q +\sqrt{(\zeta^{dis}_Q)^2 -1}).
\end{align}
The length of the connected geodesics is
\begin{align}
    \frac{d_{con}}{R}
    &= 2\log \frac{r_\infty}{R} +\log 2 + 2\,\mathrm{arcsinh}\, \lambda.
\end{align}
Thus,
\begin{equation}
    \frac{d_{dis}-d_{con}}{R}=\mathrm{arccosh}\, \zeta^{dis}_Q - 2\,\mathrm{arcsinh}\, \lambda <0,\quad \forall\lambda>0.
\end{equation}
This means the disconnected entanglement wedge is favored for any tensions (unless $\lambda\rightarrow \infty$, where $Q$ approaches the asymptotic boundary), as is illustrated in Fig.~\ref{fig:scat-brane-local}~(c).

In summary, we have seen that, in a setup with a brane, the boundary input regions $R_1$ and $R_2$ are not entangled, and as such, the bulk scattering process appears to be not realizable in the intermediate picture. 

\subsection{Violation of subadditivity on a brane}\label{sec:SA-viol-puzzle}

In this subsection, we introduce another puzzle concerning the entanglement structure for the setups with a brane.
Somewhat surprisingly, subadditivity of quantum entropy can be violated on the brane $Q$ when one computes entanglement entropies by the HRT formula. 
The violation has been already reported in~\cite{Grado-White:2020wlb}, {where null subregions on the asymptotic boundary are considered.} 
In this subsection, we present an arguably simpler example, {where all the subregions are manifestly spacelike.}

\begin{figure}
    \centering
    \includegraphics[width=0.3\linewidth]{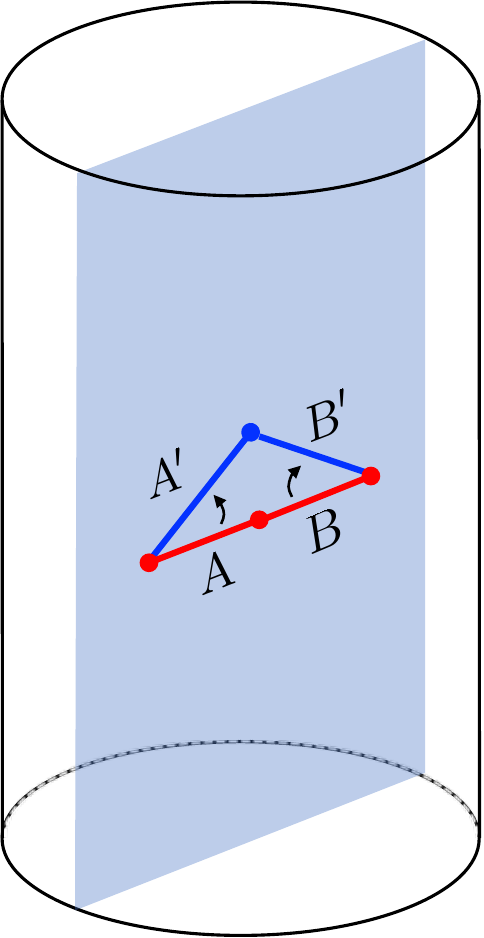}
    \caption{Geodesics connecting two points on the zero-tension brane lie on it. When $A$ and $B$ are on the same time slice, $S_A+S_B=S_{AB}$ (red). When one boosts $AB$ to $A'B'$ (blue), the proper length of each geodesic becomes shorter. Assuming the boost acts as a local unitary on $AB$, $S_{A'B'}=S_{AB}$. 
    Consequently, we have $S_{A'}+S_{B'}<S_{A'B'}$.}
    \label{fig:SA-viol-zero}
\end{figure}

For simplicity of discussion, let us consider a setup with a brane with $T\rightarrow +0$, as illustrated in Fig.~\ref{fig:SA-viol-zero}. 
The HRT formula says that entanglement entropy on a subregion $A$ on a brane is given by the geodesic length connecting the endpoints of $A$ on $Q$. 
Due to the Neumann boundary condition \eqref{eq:neumann-bc} imposed on the brane $Q$, the extrinsic curvature vanishes on $Q$ at $T=0$. Thus, any geodesics connecting two points on $Q$ must lie on $Q$ in the leading order of $G_N\rightarrow 0$. Suppose we take two adjacent subregions $A$ and $B$ on the brane as in Fig.~\ref{fig:SA-viol-zero}. Since the geodesics calculating holographic entanglement entropies lie on $Q$, we have
\begin{equation}
    S_A+S_B\approx S_{AB}.
\end{equation}
Hence, $A$ and $B$ are nearly separable.

Let us boost the interval $AB$ to $A'B'$ so that it approaches null (Fig.~\ref{fig:SA-viol-zero}). 
Since the boost must be a unitary transformation within the causal diamond, it must be implemented as a local unitary on $AB$, and leaves $S_{AB}$ unchanged. However, the geodesic lengths of $A'$ and $B'$ become small as we boost them. As such, we have
\begin{align}
S_{A'}+S_{B'} < S_A + S_B \approx S_{AB} = S_{A'B'}
\end{align}
which is a violation of subadditivity $S_{A'}+S_{B'}\ge S_{A'B'}$.

Hence, we have seen that the setup with a brane appears to violate a basic inequality that any quantum mechanical system should obey. 
As we will see later, this puzzle on subadditivity and the aforementioned scattering puzzle have the same origin and can be resolved on an equal footing by introducing the notion of induced light cones.

\section{Induced light cone on a brane}\label{sec:induced}

In order to resolve the aforementioned puzzle concerning a scattering process with a brane, it will be useful to revisit how the brane setup is related to the original AdS/CFT correspondence. 
Indeed, this seems a natural approach as, in the AdS/BCFT correspondence~\cite{Takayanagi:2011zk}, a brane can be inserted by projecting a subregion on the asymptotic boundary onto a boundary state~\cite{Numasawa:2016emc,Akal:2021dqt,Antonini:2022sfm}. 
Also, in the case of the cutoff surface, moving the boundaries radially to the bulk can be understood as coarse-graining of the dual CFT on the asymptotic boundary~\cite{Faulkner:2010jy,Heemskerk:2010hk,Balasubramanian:2012hb}.
These motivate us to return to the original dual CFT on the asymptotic boundary, where no causality puzzle is present. 

Recall that, in the original AdS/CFT correspondence, a bulk local excitation can be created by applying a local perturbation on an asymptotic boundary and then, bringing it to the bulk~\cite{Terashima:2023mcr}. 
Namely, in the dual CFT, a local excitation deep in the bulk cannot be instantaneously created and requires some preparation. 
The crux of our proposal is that the bulk local excitation at $c_1$ on $Q$ requires some preparation as well.
In particular, we propose that \emph{a local excitation on a brane $Q$, dual to a radially propagating signal in the bulk, cannot be created by a local operator on $Q$.}

\begin{figure}
    \centering
    (a)\includegraphics[width=0.3\linewidth]{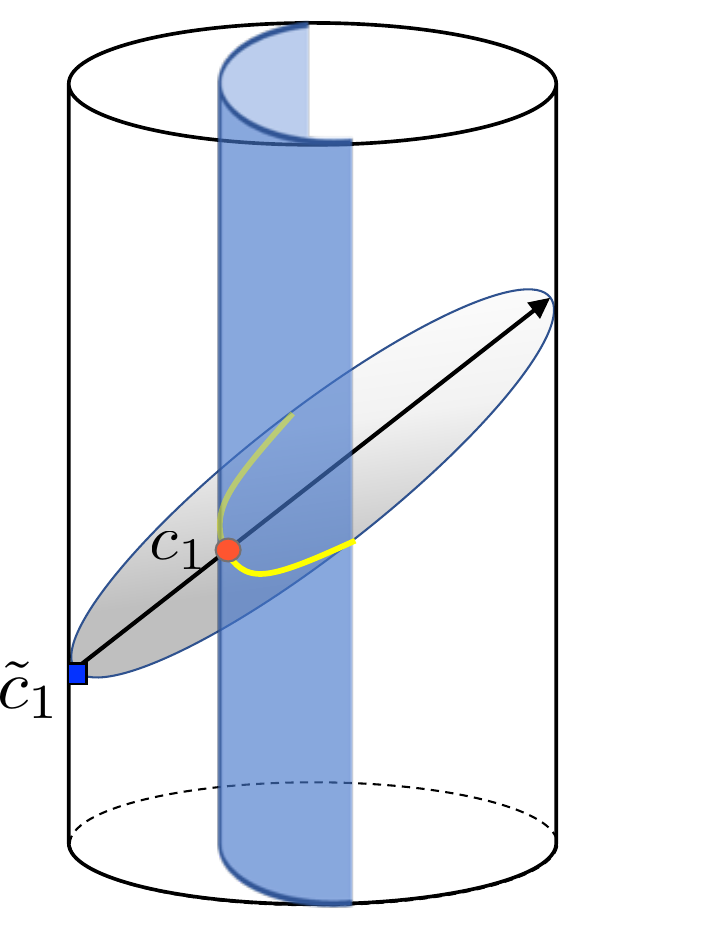}
    (b)\includegraphics[width=0.5\linewidth]{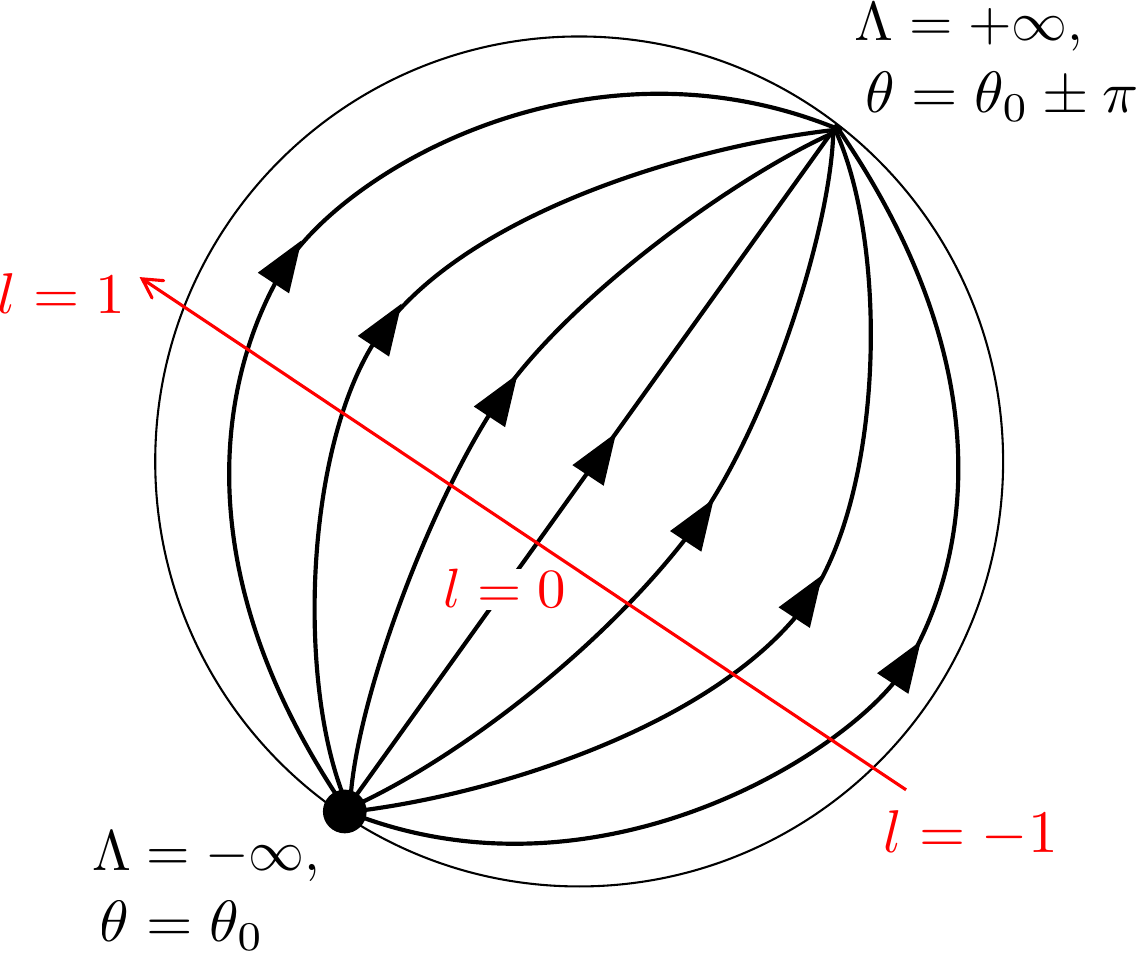}
    \caption{(a) An induced light cone (yellow curve) of an input $c_1$ (red dot). The light cone is defined as the intersection between the null surface (gray) from the local excitation $\tilde{c}_1$ (blue square) on $\tilde{\Sigma}$ and the brane $Q$ (light blue).
    (b) The top view of the bulk null surface emanating from $(t,\theta)=(-\pi/2,\theta_0)$. The codimension-one surface is parametrized by the time $\Lambda$ and the angular coordinate $l$.
    }
    \label{fig:null-surf-brane}
\end{figure}

In order to understand how a local brane excitation may be created, imagine a setup as shown in Fig.~\ref{fig:null-surf-brane}~(a) where a fictitious asymptotic boundary $\tilde{\Sigma}$ is placed behind the brane/cutoff surface $Q$.
Given the input point $c_1$ on $Q$, let us bring it back to another point $\tilde{c}_1$ on $\tilde{\Sigma}$ by considering the past light cone from $c_1$. 
The choice of $\tilde{c}_1$ depends on the direction of the light ray, and we will choose it so that it matches the direction in which $c_1$ will travel in the bulk.
We will return to this point in later sections. 

The central object of interest is an \emph{induced light cone} on $Q$. We define an induced light cone from $\tilde{c}$ as an intersection between $Q\cup\Sigma$ and the bulk light cone emanating from $\tilde{c}$ inserted at $\tilde{\Sigma}$. Formally, the future induced domain of dependence $\hat{J}\big|_Q (\tilde{c})$ is defined as (Fig.~\ref{fig:null-surf-brane} (a))
\begin{equation}
    \hat{J}^+\big|_Q (\tilde{c})\equiv J^+(\tilde{c})\cap (Q\cup \Sigma).
    \label{eq:induced-DoD}
\end{equation}
A past induced light cone can be defined similarly.

Let us derive an explicit form of the induced light cone on $Q$.
A null surface in the bulk emanating from a boundary point $(t,\theta)=(-\pi/2,\theta_0)$ to the future (or equivalently from $(\pi/2,\theta_0 \pm\pi)$ to the past) is parametrized by $\Lambda=\tan t$ and $l\in (-1,1)$~\cite{May:2019yxi}:
\begin{equation}
    \begin{split}
        \frac{r^2+R^2}{R^2} &= \frac{1+\Lambda^2}{1-l^2}, \\
        -\cot(\theta-\theta_0) &= \frac{\Lambda}{l}
    \end{split}
    \label{eq:null-traj}
\end{equation}
as shown in Fig.~\ref{fig:null-surf-brane}~(b).
To find the intersection with the ETW brane $Q$, we substitute $r\sin\theta=-\lambda R$ (Eq.~\eqref{eq:brane-global}) into the above equations:
\begin{equation}
    \begin{split}
        \frac{\lambda^2}{\sin^2\theta}+1 &= \frac{1+\Lambda^2}{1-l^2}, \\
        -\cot(\theta-\theta_0) &= \frac{\Lambda}{l}.
    \end{split}
\end{equation}
Eliminating $l$, we obtain
\begin{align}
    \Lambda^2=\tan^2 t
    &=\frac{\lambda^2 (1+ \tan\theta \tan\theta_0)^2}{\tan^2\theta\sec^2\theta_0 +\lambda^2 (\tan\theta-\tan\theta_0)^2} .
\end{align}
By using \eqref{eq:eta-theta}, we can express $\tan^2 t$ in terms of $\eta$ as
\begin{align}
    \tan^2 t &= \frac{\qty(\sqrt{1+\lambda^2}+\abs{\lambda}\tan\theta_0 \tan\eta)^2}{\sec^2\theta_0 \tan^2\eta + \qty(\abs{\lambda} \tan\eta - \sqrt{1+\lambda^2}\tan\theta_0)^2}.
    \label{eq:t-eta-squared}
\end{align}
This gives the explicit trajectory of the future/past induced light cones on $Q$ from a point $(t,\theta)=(\pm\pi/2,\theta_0)$.
Note that, when $\theta_0=0,\pi$, their induced light cones are given by $t=\pm(\abs{\eta}-\pi/2)$, the same as the usual light cone.

\begin{figure}
    \centering
    \includegraphics[width=0.7\linewidth]{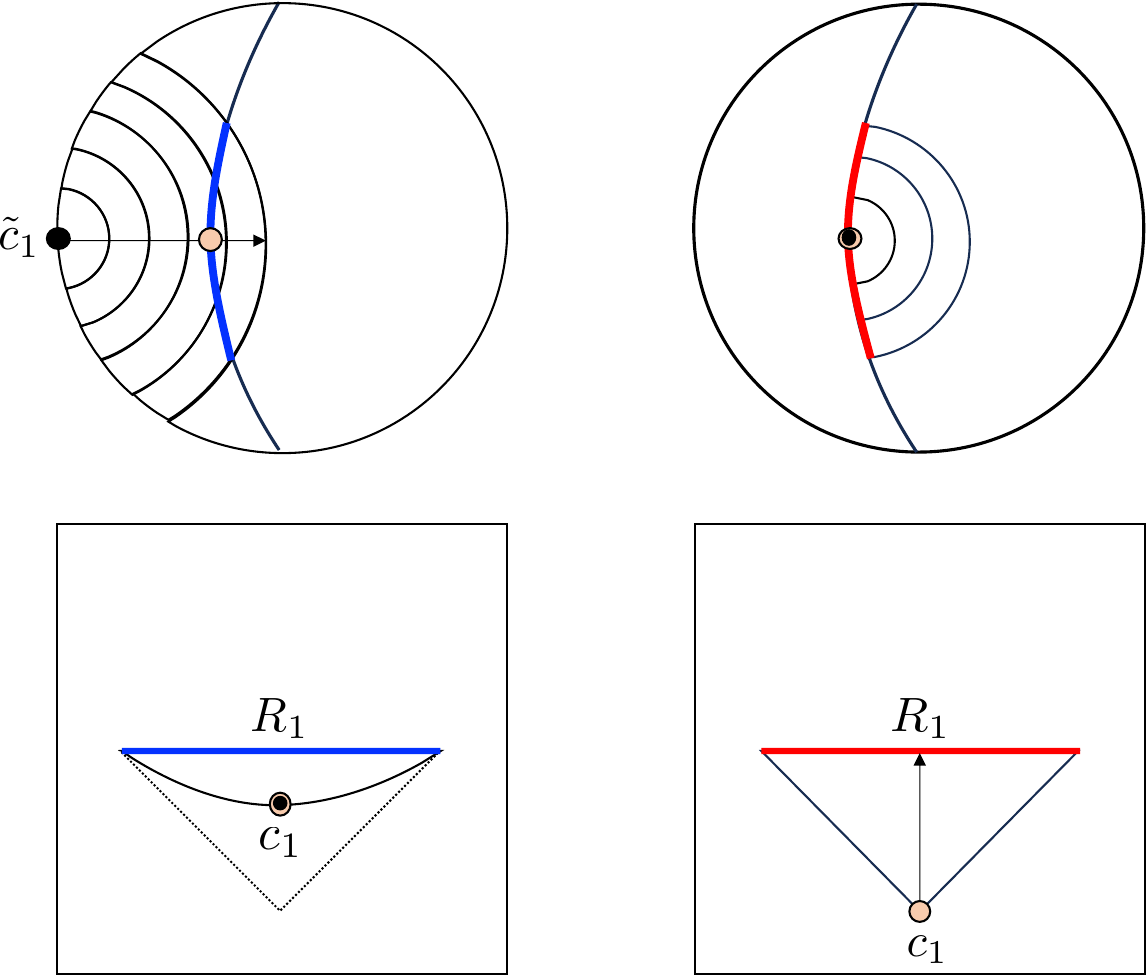}
    \caption{
    A radially propagating excitation v.s. trapped excitation.
    Left: A radially propagating excitation at $c_1$ was prepared as a local excitation at $\tilde{c}_1$ on the fictitious boundary  $\tilde{\Sigma}$ followed by its subsequent time evolution. 
    The intersection of the bulk null surface with $Q$, illustrated in blue, gives rise to the induced light cone of a radially propagating excitation.
    Right: A local excitation, prepared on $Q$ by a local operator obeys the standard causality. 
    Such an excitation is trapped on $Q$ and does not propagate toward the bulk.
   }
    \label{fig:loc-vs-nonloc}
\end{figure}

The difference between the standard light cone and the induced light cone from $c_1$ is shown in Fig.~\ref{fig:loc-vs-nonloc}. Notice that the induced light cone is not null on the brane, and thus seemingly allows superluminal signaling.
Our proposal is that it is the induced light cone with apparent superluminal signaling which is responsible for the bulk direct scattering.
Namely, preparation for creating a radially propagating excitation at $c_1$ should start earlier than $c_{1}$ as depicted in Fig.~\ref{fig:loc-vs-nonloc}.\footnote{This should remind readers of how an apparent superluminal propagation occurs by careful initial preparation. A useful analogy is to imagine local agents placed on a one-dimensional line where they are instructed to raise flags at some specific time. Then, one can program them as if the wave of raising flags propagates superluminally. This is an apparent propagation and does not transmit any information.
}
In the rest of the paper, we will demonstrate that this approach resolves the scattering puzzle, as well as several other causality and entanglement puzzles on a brane, on an equal footing. 

\section{Connected wedge theorem from induced light cones}\label{sec:CWT}

In this section, we argue that induced light cones resolve the bulk scattering puzzle. We begin by demonstrating this for the setup considered in Section~\ref{sec:scat-brane-local} via explicit calculations.
Namely, we will see that, in this case, the geodesic lengths for connected and disconnected wedges precisely match, and thus the setup is exactly at the transition. We then provide a generic proof by extending the connected wedge theorem for the setup with an arbitrary cutoff surface.
The proof extends a geometric argument from~\cite{May:2019odp} and applies to the cases where multiple input and output points are located on the cutoff surface. 

\subsection{An explicit example}

Consider the same setup provided in Section~\ref{sec:scat-brane-local} where the input $c_2$ on the asymptotic boundary $\Sigma$ is given by $(t,\theta)=(-\pi/2,\pi/2)$, the outputs $r_1,r_2$ are given by $(\pi/2,0)$ and $(\pi/2,\pi)$, and the input $c_1$ on the brane $Q$ is given by $(t,\eta)=(-\arctan\lambda,-\pi/2)$. The input point $\tilde{c_1}$ on the fictitious boundary $\tilde{\Sigma}$ is given by $(t,\theta)=(-\pi/2,-\pi/2)$. 

The essential difference is that the input region  $D(R_i)$ needs to be determined from the future induced light cone $\partial \hat{J}^+\big|_Q (\tilde{c}_i)$ of $c_i$, 
the past induced light cone $\partial \hat{J}^-\big|_Q (\tilde{r}_1)$ of $r_1$, and the past induced light cone $\partial \hat{J}^-\big|_Q (\tilde{r}_2)$ of $r_2$ instead of~\eqref{eq:DoD-R2}. 
In the current case, $R_2$ is the same as the previous consideration; $t=-\pi/4$ and $\theta=[\pi/4,3\pi/4]$ lying on $\Sigma$. Meanwhile, 
$R_1$ is defined from the intersection of \emph{induced} domains of dependence~\eqref{eq:induced-DoD}:
\begin{equation}
    D\big|_Q (R_1)\equiv J^+(\tilde{c_1})\cap J^-(r_1) \cap J^-(r_2) \cap (Q\cup\Sigma).
\end{equation}
Here, $\partial R_1$ is given by the intersection of
\begin{align*}
    \partial \hat{J}^- (r_1):&\quad t=\eta +\frac{\pi}{2}, \\
    \partial \hat{J}^- (r_2):&\quad t=-\eta -\frac{\pi}{2}, 
\end{align*}
and
\begin{equation}
    \partial \hat{J}^+\big|_Q (\tilde{c}_1): \quad  
    t={\arctan\qty(\frac{\lambda \tan\eta}{\sqrt{1+\lambda^2+\tan^2\eta}})}.
\end{equation}
By solving this, we obtain
\begin{equation}
    R_1: \quad t=-\frac{\pi}{2}+\eta_1,\quad \eta=\qty[-\pi+\eta_1, -\eta_1],
\end{equation}
where $\eta_1 \in [0,\frac{\pi}{2})$ is taken such that $\tan\eta_1 = \frac{\sqrt{1+\lambda^2}}{\lambda}$.

Once $R_1$ and $R_2$ are specified, their holographic entropies can be calculated. For a disconnected entanglement wedge, the length of the geodesic for $R_2$ is calculated from \eqref{eq:discon-geo-R2}. For $R_1$, we have
\begin{equation}
    \zeta_Q^{dis} = 2(1+\lambda^2) \cot^2\eta_1 +1 = 1+2\lambda^2
\end{equation}
from \eqref{eq:zeta-discon-bdy}. From these, we can calculate the length of the disconnected geodesics from \eqref{eq:geodesic-gen} and \eqref{eq:geodesic-bdy-lim}:
\begin{equation}
    \frac{d_{dis}}{R}=2\log \frac{r_\infty}{R} + \log 2 + \log\qty(1+2\lambda^2 +2\lambda \sqrt{1+\lambda^2}).
    \label{eq:induce-dis-geo}
\end{equation}
For a connected geodesic, we have
\begin{equation}
    \zeta_{\Sigma Q}^{con}= \frac{1}{\sqrt{2}}\frac{r_\infty}{R}\qty(\sqrt{1+\lambda^2}+\lambda)
\end{equation}
from \eqref{eq:zeta-discon-Q}. This gives the length of the geodesics connecting $\partial R_1$ and $\partial R_2$:
\begin{equation}
    \frac{d_{con}}{R}= 2 \qty[\log 2 + \log\qty(\frac{1}{\sqrt{2}}\frac{r_\infty}{R}) + \log \qty(\sqrt{1+\lambda^2}+\lambda)].
    \label{eq:induce-con-geo}
\end{equation}
From \eqref{eq:induce-dis-geo} and \eqref{eq:induce-con-geo}, their difference is given by
\begin{equation}
    \frac{d_{dis}-d_{con}}{R}= \log\frac{1+2\lambda^2 +2\lambda \sqrt{1+\lambda^2}}{(\sqrt{1+\lambda^2}+\lambda)^2} = 0. \label{eq:match}
\end{equation}

We have found that not only is a connected entanglement wedge preferred, $d_{dis}$ and $d_{con}$ precisely match in this case. 
Note that arguments based on holographic tasks dictate the connected wedge only ($d_{dis}\geq d_{con}$) and do not necessarily demand $d_{dis}=d_{con}$ when the bulk scattering region $P$ consists of a point. 
Yet, this ``$d_{dis}=d_{con}$ for $P$ being a point'' is actually an interesting property often observed in many examples of the bulk scattering in the original AdS/CFT setup~\cite{May:2019yxi,May:2019odp,May:2021nrl,May:2021zyu,May:2022clu}.
Hence, \eqref{eq:match} may be taken as additional evidence that the induced light cone is indeed the right object to consider.

Finally, we remark on the arbitrariness concerning the choice of $\tilde{c}_1$. In the present setup, our choice of $\tilde{c}_1$, as a direct past of $c_1$'s propagation trajectory in the bulk, seems natural. 
It is however, in principle, possible to prepare an excitation at $c_1$ from other points on the fictitious boundary $\tilde{\Sigma}$ as long as they are in the past light cone of $c_1$.
It turns out that the connected wedge holds for any choice of $\tilde{c}_1$ as long as it is causally connected to $c_1$.
This fact is proven in the next subsection from a geometric argument.

\subsection{Geometric proof}\label{sec:geom-proof-CWthm}

We have seen that defining input regions $R_{1}, R_{2}$ via induced light cones may lead to the connected entanglement wedge.
In this subsection, we prove the connected wedge theorem for setups with arbitrary cutoff surfaces including inhomogeneous ones.

The original connected wedge theorem is the following statement:

\begin{itemize}[leftmargin=*]
\item[] \textbf{Connected wedge theorem:} Suppose that $c_{1}, c_{2}\rightarrow r_{1},r_{2}$ scattering is possible in the bulk, but is not possible on the boundary. Then, the entanglement wedge for $R_{1}\cup R_{2}$ is connected.
\end{itemize}

\begin{figure}
    \centering
    (a)
    \includegraphics[width=0.45\textwidth]{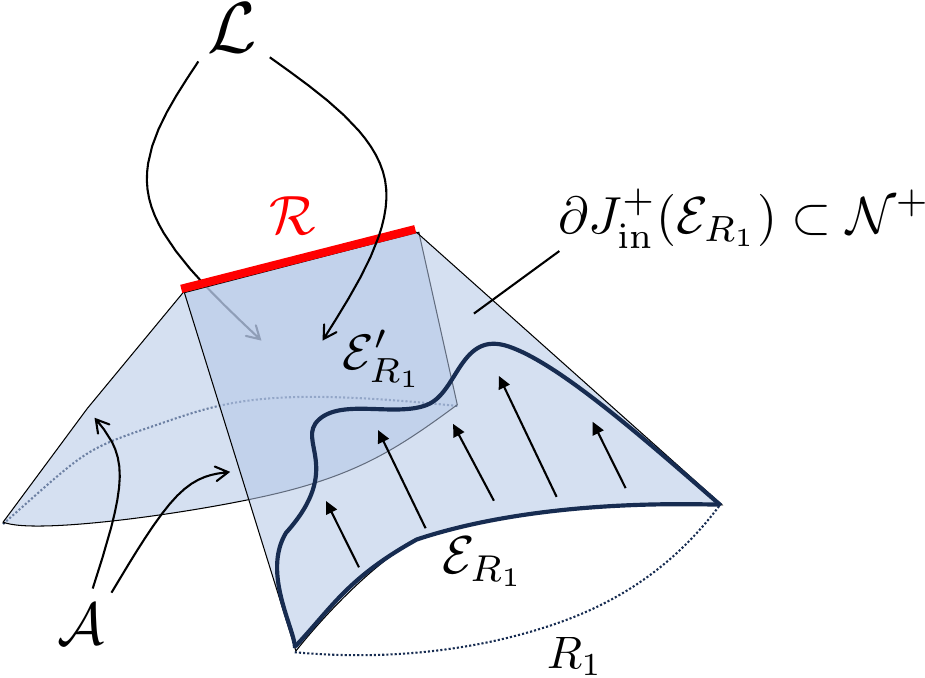}
    \
    (b)
    \includegraphics[width=0.4\textwidth]{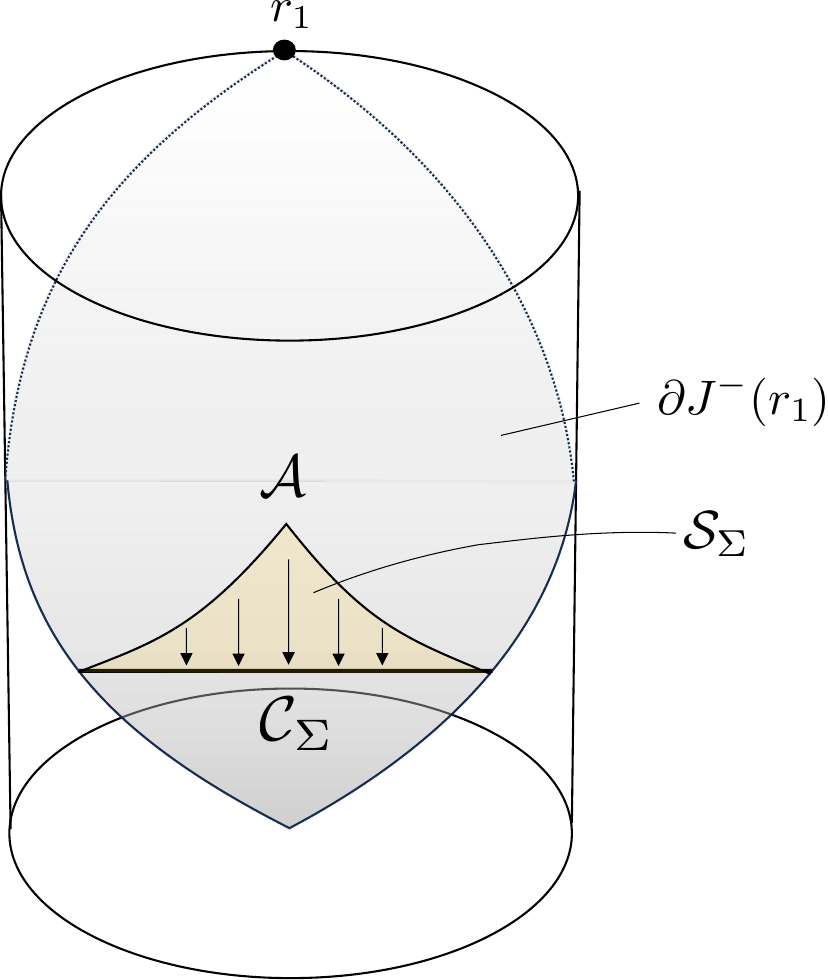}
    \caption{The null membrane $\mathcal{N}_\Sigma$ consists of the lift $\mathcal{L}$ and the slope $\mathcal{S}_\Sigma$. (a) The lift $\mathcal{L}$ is shown as  a blue sheet. It is part of $\partial J_{\mathrm{in}}^+$, namely, lightsheets extending inward from both $\mathcal{E}_{R_1}$ and $\mathcal{E}_{R_2}$. The boundary of two null planes constituting~$\mathcal{L}$ consists of the ridge $\mathcal{R}$ and the seam $\mathcal{A}$. Due to non-positive expansion, $\mathrm{area}(\mathcal{E}'_{R_1})\le \nobreak\mathrm{area}(\mathcal{E}_{R_1})$. 
    (b) The slope $\mathcal{S}_\Sigma$ is shown as a dark yellow surface. It is on the future causal horizon $\partial J^-(r_1)\cup \partial J^-(r_2)$. (Only one side is shown.) Its boundary is the seam $\mathcal{A}$ and a connected surface $\mathcal{C}_\Sigma$. Due to non-negative expansion with respect to past-direct null geodesics, $\mathrm{area}(\mathcal{A})\ge \mathrm{area}(\mathcal{C}_{\Sigma})$.}
    \label{fig:lift-slope}
\end{figure}

Let us briefly recall an important underlying assumption behind the geometric proof in~\cite{May:2019odp}. The proof of the connected wedge theorem relies on two statements that follow from the \emph{focusing theorem}. The first statement concerns the lightsheets of an extremal surface $\mathcal{E}$, which can be defined as the boundary of the future (or past) of $\mathcal{E}$; $\mathcal{N}^{\pm } = \partial J^{\pm}(\mathcal{E})$~\cite{Bousso:1999xy}. $\mathcal{N}^{\pm}$ has everywhere non-positive expansion. The second statement concerns the causal horizon, which is the boundary of the past of a point $P$ at infinity. Specifically, for the future causal horizon $\partial J^{-}(P)$ generated by past-directed null geodesics, its expansion must be non-positive with respect to past-directed generators. This non-positivity/negativity of expansion results in the non-increasing/decreasing area, respectively. See Fig.~\ref{fig:lift-slope} for examples.

We begin with a quick review of the geometric argument. Suppose that the entanglement wedge for $R_{1}\cup R_{2}$ is disconnected; $\mathcal{E}_{R_{1}\cup R_{2}} =\mathcal{E}_{R_{1}} \cup \mathcal{E}_{R_{2}}$. 
The key idea is to find a connected surface $\mathcal{C}_{\Sigma}$ whose area is smaller than that of  $\mathcal{E}_{R_{1}} \cup \mathcal{E}_{R_{2}}$, for any complete achronal slice $\Sigma$ containing $\mathcal{E}_{R_{1}} \cup \mathcal{E}_{R_{2}}$. This would contradict with that the HRT surface must be a global minimum on some complete achronal slice, as implied by the maximin procedure~\cite{Wall:2012uf}. Hence, the entanglement wedge for $R_{1}\cup R_{2}$ must be connected. 

\begin{figure}
\centering
\includegraphics[width=0.6\textwidth]{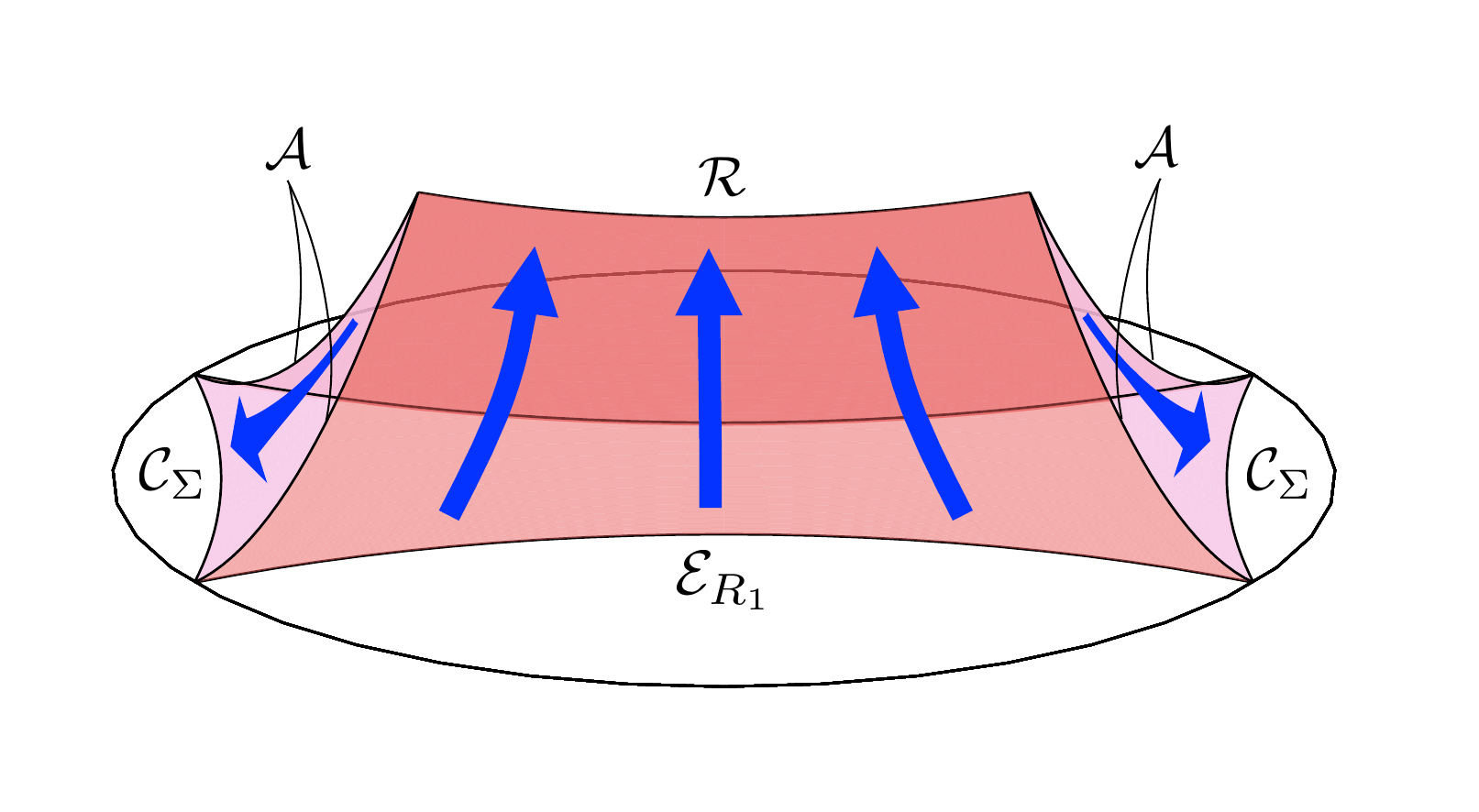}
\caption{Lightsheets from extremal surfaces $\mathcal{E}_{R_{1}},\mathcal{E}_{R_{2}}$ meet at the ridge $\mathcal{R}$ and leave the seam $\mathcal{A}$. The seam $\mathcal{A}$ is a part of the causal horizons from $r_{1},r_{2}$, finding a connected surface $\mathcal{C}_{\Sigma}$ on $\Sigma$. 
Deforming $\mathcal{E}_{R_1}\cup \mathcal{E}_{R_2}$ along the lift, then along the slope, leads to $\mathrm{area}(\mathcal{E}_{R_1}\cup\nobreak \mathcal{E}_{R_2})\ge \nobreak\mathrm{area}(\mathcal{C}_\Sigma)$, as indicated by blue arrows. The figure is taken from~\cite{May:2019odp} and modified with permission.
}
\label{fig-notation}
\end{figure}

To find such a connected surface $\mathcal{C}_{\Sigma}$, we begin by constructing a certain codimension\nobreakdash-$1$ null surface, called the null membrane $\mathcal{N}_{\Sigma}$. For any complete achronal slice $\Sigma$ containing $\mathcal{E}_{R_{1}} \cup \mathcal{E}_{R_{2}}$, $\mathcal{N}_{\Sigma}$ consists of two distinct pieces. The first piece, called the lift $\mathcal{L}$, is defined by 
\begin{align}
\mathcal{L} = \partial J^{+}_{\mathrm{in}}( \mathcal{E}_{R_{1}} \cup \mathcal{E}_{R_{2}} ) \cap J^{-}(r_{1})\cap J^{-}(r_{2}),
\end{align}
where $\partial J^+_{\mathrm{in}}\subset \partial J^+$ consists of lightsheets containing null geodesics initially moving away from the asymptotic boundary.
The lift $\mathcal{L}$ has a special structure with a spacelike ridge $\mathcal{R}$ where the lightsheets from $\mathcal{E}_{R_{1}}$ and $\mathcal{E}_{R_{2}}$ meet (Fig.~\ref{fig:lift-slope} (a)). The fact that the ridge $\mathcal{R}$ is nonempty follows from the assumption that the bulk scattering region $J_{12\rightarrow 12}$ is nonempty and the fact that the HRT surface of $R_{j}$ must always lie spacelike outside the causal wedge of $R_{j}$~\cite{Headrick:2014cta,Wall:2012uf,Engelhardt:2014gca}. The second piece, called the slope $\mathcal{S}_{\Sigma}$, is defined by 
\begin{align}
\mathcal{S}_{\Sigma} = \partial [ J^{-}(r_{1}) \cap J^{-}(r_{2}) ] \cap J^{-}[ \partial J^{+}_{\mathrm{in}}( \mathcal{E}_{R_{1}} \cup \mathcal{E}_{R_{2}} )] \cap J^{+}(\Sigma).
\end{align} 
It is worth emphasizing that the slope $\mathcal{S}_{\Sigma}$ is part of the causal horizon with respect to $r_{1}$ and $r_{2}$. The geometric structure of the slope $\mathcal{S}_{\Sigma}$ is shown in Fig.~\ref{fig:lift-slope} (b). 
The null membrane $\mathcal{N}_{\Sigma}$ is a union of these two pieces; $\mathcal{N}_{\Sigma} = \mathcal{L}\cup \mathcal{S}_{\Sigma}$.  

We now show that the past boundary of the slope $\mathcal{S}_{\Sigma}$, denoted by $\mathcal{C}_{\Sigma}$, has an area smaller than that of  $\mathcal{E}_{R_{1}} \cup \mathcal{E}_{R_{2}}$. Let us denote spacelike seams where the lift and slope meet by $\mathcal{A}$ (Fig.~\ref{fig-notation}). Note that the lift $\mathcal{L}$ is part of the future lightsheets of extremal surfaces. So, it has non-positive expansion, implying 
\begin{align}
\mathrm{area}(\mathcal{E}_{R_{1}} \cup \mathcal{E}_{R_{2}}) \geq \mathrm{area}(\mathcal{A}) + 2\,\mathrm{area}(\mathcal{R}) \geq \mathrm{area}(\mathcal{A}).
\end{align}
Similarly, note that the slope is a part of the causal horizons of points $r_{1},r_{2}$ at infinity. So, it also has non-positive expansion, implying 
\begin{align}
\mathrm{area}(\mathcal{A}) \geq \mathrm{area}(\mathcal{C}_{\Sigma}).
\end{align}
Hence, we arrive at 
\begin{align}
\mathrm{area}(\mathcal{E}_{R_{1}} \cup \mathcal{E}_{R_{2}})  \geq \mathrm{area}(\mathcal{C}_{\Sigma}).
\end{align}

\begin{figure}
\centering
\includegraphics[width=0.3\textwidth]{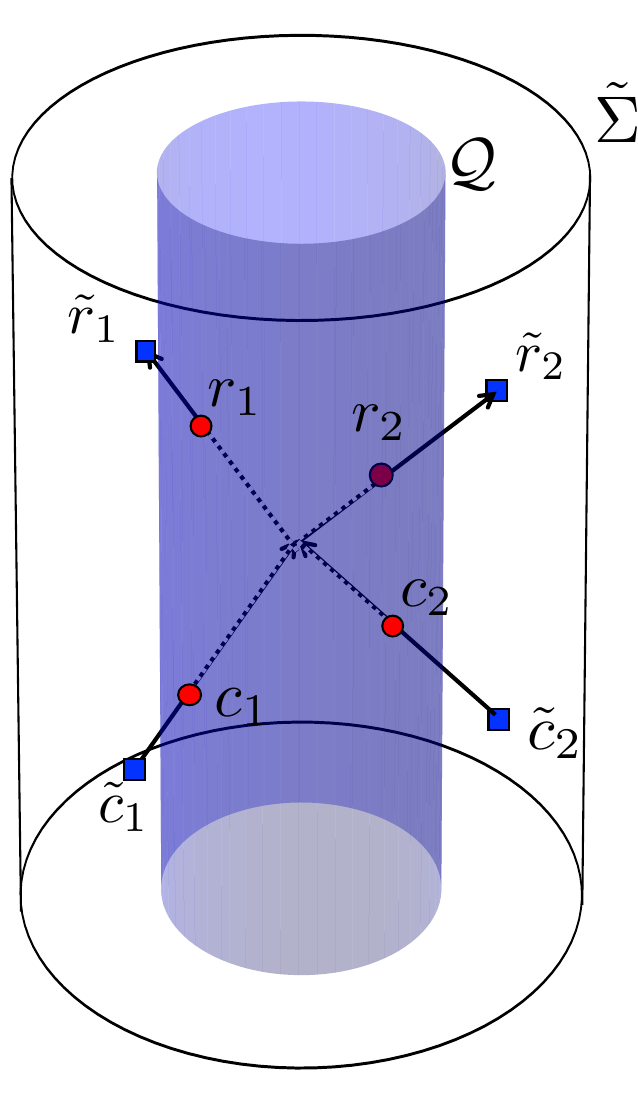}
\caption{ 
The input/output points $c_1,c_2,r_1,r_2$ are placed on the cutoff surface $\mathcal{Q}$, denoted by the blue cylinder. 
In this example, it is placed on a constant radius.
Blue squares denote the causally past/future points $\tilde{c}_1,\tilde{c}_2,\tilde{r}_1,\tilde{r}_2$ on $\tilde{\Sigma}$ of inputs/outputs, respectively.
}
\label{fig:cutoff-fict}
\end{figure}

Having reviewed the proof technique, let us generalize the connected wedge theorem for setups with an arbitrary cutoff surface $\mathcal{Q}$. Let $\{ \tilde{c}_{1},\tilde{c}_{2},\tilde{r}_{1},\tilde{r}_{2} \}$ be arbitrary points on the boundary of an asymptotically AdS spacetime which are causally connected to $\{c_{1},c_{2},r_{1},r_{2}\}$ on the cutoff surface $\mathcal{Q}$. 
Namely, we assume 
\begin{align}
\tilde{c}_{1} \in J^{-}(c_1), \quad \tilde{c}_{2} \in J^{-}(c_2),\quad \tilde{r}_{1} \in J^{+}(r_1), \quad \tilde{r}_{2} \in J^{+}(r_2).
\end{align}
Note that they can be located on either the true or fictitious asymptotic boundary $\Sigma$ or $\tilde{\Sigma}$, where $\tilde{\Sigma}$ denotes the asymptotic boundary of AdS behind $\mathcal{Q}$. When the cutoff surface does not anchor on the boundary, we have $\Sigma=\emptyset$. Fig.~\ref{fig:cutoff-fict} shows one such example.

Note that $\tilde{c}_{1},\tilde{c}_{2}\rightarrow \tilde{r}_{1},\tilde{r}_{2}$ scattering is possible in the bulk since $c_{1},c_{2}\rightarrow r_{1},r_{2}$ scattering is possible. Here we suppose that the boundary scattering is not possible for $\{ \tilde{c}_{1},\tilde{c}_{2},\tilde{r}_{1},\tilde{r}_{2} \}$ on the asymptotic boundary $\Sigma\cup\tilde{\Sigma}$. We further assume that $\tilde{c}_{1},\tilde{c}_{2}\rightarrow \tilde{r}_{1},\tilde{r}_{2}$ scattering is not possible outside the region surrounded by $\mathcal{Q}$ so that $R_{1}$ and $R_{2}$ are spacelike separated.

The statement of the connected wedge theorem with a cutoff surface is as follows.
\begin{itemize}[leftmargin=*]
\item[] \textbf{Generalized connected wedge theorem:} Let $R_{1}, R_{2}$ be regions on the surface $\mathcal{Q}$ defined by
\begin{align}
R_{j} = J^{+}(\tilde{c}_{j}) \cap J^{-}(\tilde{r}_{1}) \cap J^{-}(\tilde{r}_{2}) \cap (\mathcal{Q}\cup \Sigma).
\end{align}
Then $R_{1}$ and $R_{2}$ are spacelike separated and the entanglement wedge for $R_{1}\cup R_{2}$ is connected.
\end{itemize}

Below, for simplicity of discussion, we will employ the simplifying assumption.
\begin{itemize}[leftmargin=*]
\item[] \textbf{Assumption:} $R_{1}$ and $R_{2}$ consist of connected regions.
\end{itemize}

While this was indeed the case in the AdS/CFT, this may no longer be true for setups with $\mathcal{Q}$. For instance, by placing some obstacle, such as a dustball, in front of $\tilde{c}_{1}$, we may have two spacelike separated regions in $R_{1}$. Another possibility of such arises when $\mathcal{Q}$ is not convex. 

Below, we present a sketch of the proof.
The proof mostly parallels with that of the original connected wedge theorem. 
Here, it is worth discussing what would go wrong if we used standard light cones instead of induced ones. 
Crucial steps in the original proof are the repeated uses of the focusing theorem to find a connected surface $\mathcal{C}_{\Sigma}$ with a smaller area. 
In order to use the focusing arguments, however, one needs to consider a causal horizon of a point at \emph{infinity} (or, on the asymptotic boundary).
In the setup with $\mathcal{Q}$, the output points $r_1$ and $r_2$ are no longer at infinity, and thus the focusing theorem does not apply.

We first observe that the null membrane $\mathcal{N}_{\Sigma}$ in the presence of the cutoff surface $\mathcal{Q}$ has the same structure as Fig.~\ref{fig-notation}. This follows from the fact that the entanglement wedge $\mathcal{E}_{R_{1}}$ lies spacelike outside the causal wedge of $\tilde{c}_{1}$~\cite{Headrick:2014cta,Wall:2012uf,Engelhardt:2014gca}. 
We then use the focusing theorem repeatedly. 
Noting that the induced light cones were constructed from points on the asymptotic boundary. 
Hence, the area reduction argument remains valid and one is able to construct a connected surface to derive a contradiction.

In summary, we presented a generalization of the connected wedge theorem for an arbitrary cutoff surface $\mathcal{Q}$. 
We conclude this section by noting that the above statement holds for arbitrary choices of $\{ \tilde{c}_{1},\tilde{c}_{2},\tilde{r}_{1},\tilde{r}_{2} \}$ that are causally connected to $\{c_{1},c_{2},r_{1},r_{2}\}$. 
While the constructions of $R_1$ and $R_2$ depend on the choices  $\{ \tilde{c}_{1},\tilde{c}_{2},\tilde{r}_{1},\tilde{r}_{2} \}$, they are always connected. 
This is how our theorem avoids the ambiguity associated with the choice of  $\{ \tilde{c}_{1},\tilde{c}_{2},\tilde{r}_{1},\tilde{r}_{2} \}$, as mentioned at the end of the previous subsection. 

{We also note that the fictitious spacetime behind $Q$ or $\mathcal{Q}$ need not be the vacuum AdS spacetime. Since the above proof only uses the focusing theorem and the causal wedge inclusion, we expect that our generalization of the connected wedge theorem remains true for any extension to asymptotic infinity.}

\section{Subadditivity from induced light cones}\label{sec:SA-ind}
In this section, we show that our induced light cone proposal resolves the subadditivity violation puzzle from Section~\ref{sec:SA-viol-puzzle}. We begin by demonstrating this for the setup considered in Section~\ref{sec:SA-viol-puzzle} via explicit calculations.
Namely, we will see that the regime where subadditivity holds exactly matches with the induced causal diamond. We then provide a generic geometric proof that subadditivity of holographic entanglement entropy remains valid within the induced causal diamond. 

\subsection{An explicit example}

\begin{figure}
    \centering
    \includegraphics[width=0.5\linewidth]{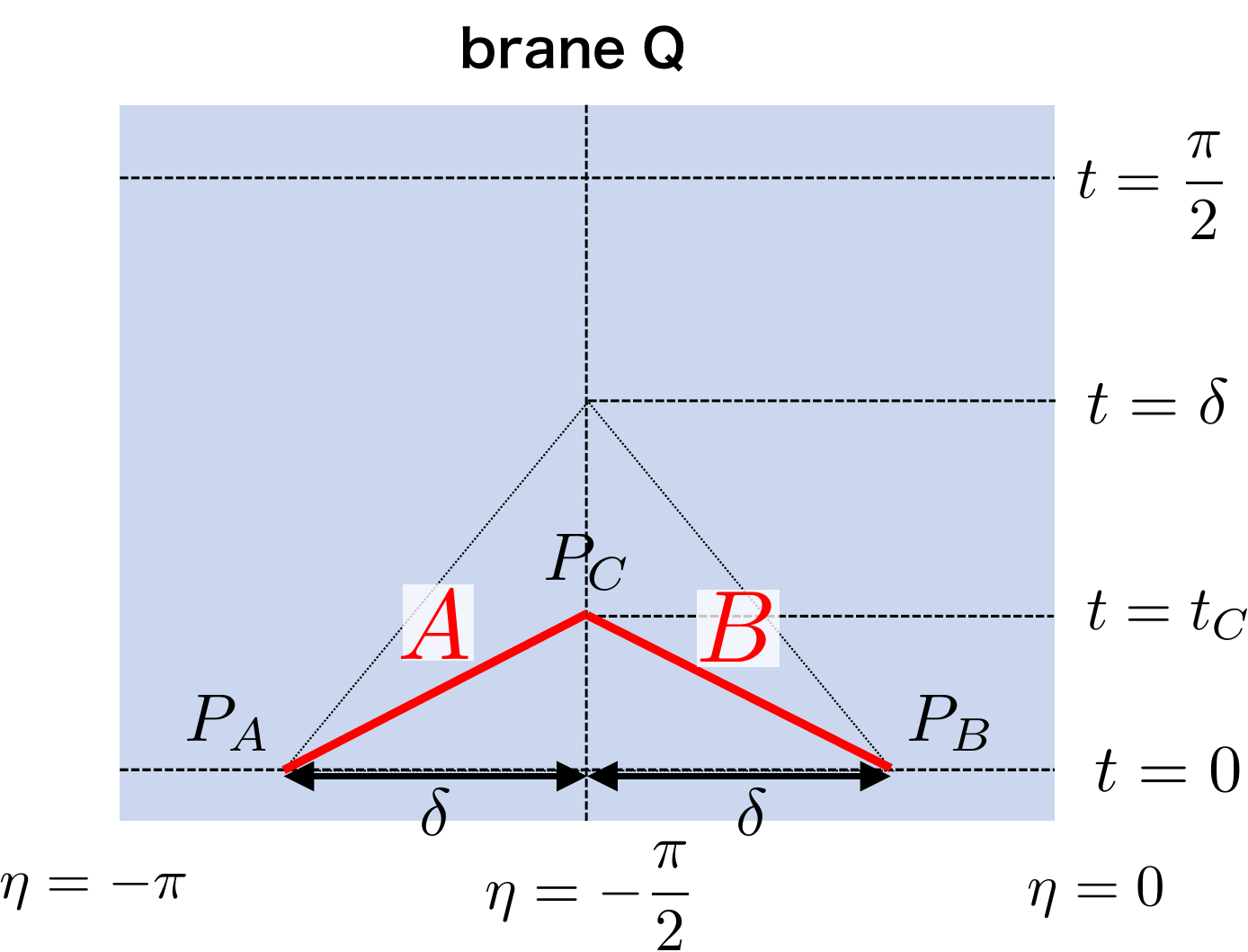}
    \caption{Subadditivity on the brane is examined by changing $t_C$ from 0 to $\delta$, where $0<\delta\le \pi/2$.}
    \label{fig:SA-check}
\end{figure}

Consider two adjacent, achronal intervals $A:[P_A,P_C]$ and $B:[P_C,P_B]$ on $Q$ placed symmetrically around $\eta = - \pi/2$, as illustrated in Fig.~\ref{fig:SA-check}. The coordinates of $P_A$ and $P_B$ are parametrized as $(t,\eta)=(0,-\pi/2-\delta),(0,-\pi/2+\delta)$ while $P_C$ is located at
\begin{equation}
    (t,\eta)=\qty(t_C,-\frac{\pi}{2}),\quad 0\le t_C \le \delta \le \frac{\pi}{2}.
    \label{eq:constraint-tC}
\end{equation}

As we have seen in Section~\ref{sec:SA-viol-puzzle}, subadditivity of holographic entanglement entropy can be violated as we increase $t_C$ beyond some critical value. 
Holographic entanglement entropy for $AB$ is calculated from \eqref{eq:zeta-discon-Q}:
\begin{equation}
    \zeta_{AB} = 2(1+\lambda^2)\tan^2\delta+1.
\end{equation}
The corresponding geodesic length is
\begin{equation}
    d_{AB}=R\, \mathrm{arccosh}\, \zeta_{AB}.
\end{equation}
Holographic entanglement entropy for $A$ or $B$ with finite $t_C<\delta$ is calculated from
\begin{equation}
    \zeta_{A}=\zeta_B = (1+\lambda^2) \frac{\cos t_C}{\cos\delta} -\lambda^2.
\end{equation}
Note that $\zeta_A >0$ since $\cos t_C > \cos\delta$ from \eqref{eq:constraint-tC}.
The geodesic length for $A$ and $B$ is
\begin{equation}
    d_{A}=d_B=R\, \mathrm{arccosh}\, \zeta_{A}.
\end{equation}
Hence, subadditivity of holographic entanglement entropy on $Q$ constrains $t_C$ as follows:
\begin{align}
    d_A+d_B \ge d_{AB} &\Leftrightarrow 2 \, \mathrm{arccosh} \zeta_A \ge \mathrm{arccosh}\, \zeta_{AB} \nonumber\\
    &\Leftrightarrow \cos t_C \ge \frac{\cos\delta}{1+\lambda^2}\qty(\lambda^2+\sqrt{(1+\lambda^2)\tan^2\delta +1}).
    \label{eq:SA-region}
\end{align}

We now demonstrate that this condition ensures $P_C$ lies within the induced causal diamond. The induced causal diamond can be constructed from the past and future induced light cones so that their intersections match with $\partial(AB)$. 
A general induced light cone is given by \eqref{eq:t-eta-squared} up to the shift in $t$. Let us denote the shift by $\pm t_0$. Since $AB$ is symmetric along $\eta=-\pi/2$, we take $\theta_0=-\pi/2$. Then, the edge of the induced causal diamond is given by
\begin{equation}
    t \pm t_0 = \pm \arctan\qty[\abs{\frac{\lambda \tan\eta}{\sqrt{1+\lambda^2+\tan^2\eta}}}]
    \in\left(-\frac{\pi}{2},\frac{\pi}{2}\right].
\end{equation}
The $+$ sign corresponds to the upper edge and the $-$ sign corresponds to the lower edge of the induced causal diamond.
By equating their intersections with $\partial (AB):(t,\eta)=(0,-\pi/2\pm\delta)$, $t_0$ is fixed:
\begin{equation}
    t_0 = \arctan\qty[\frac{\lambda}{\sqrt{(1+\lambda^2)\tan^2 \delta +1}}]
    \in\qty[0,\frac{\pi}{2}].
\end{equation}
For $P_C$ to be within the induced causal diamond, we need
\begin{equation}
    0\le t_C \le t(\eta=-\pi/2), 
    \label{eq:constraint-tC2}
\end{equation}
where
\begin{equation}
    t(\eta=-\pi/2)=\arctan\lambda - t_0
\end{equation}
for finite $\lambda$.
From \eqref{eq:constraint-tC2}, we find
\begin{align}
    \cos t_C &\ge \cos t(\eta=-\pi/2) \nonumber\\
    & = \frac{\cos\delta}{1+\lambda^2}\qty(\lambda^2+\sqrt{(1+\lambda^2)\tan^2\delta +1}).
\end{align}
This is exactly the regime \eqref{eq:SA-region}, where subadditivity of holographic entanglement entropy holds. 

In summary, we have demonstrated that subadditivity of holographic entanglement entropy holds for an interval $AB$ when we restrict our focus to a particular induced caudal diamond that can be constructed by shifting \eqref{eq:t-eta-squared}. 
This induced diamond originates from some points at $\eta=-\frac{\pi}{2}$ on the (fictitious) asymptotic boundary. 
Note, however, that the construction of the induced causal diamond depends on the choice of these points on the asymptotic boundary and thus is not unique.
In the next subsection, we present general proof that subadditivity holds on arbitrary induced causal diamonds.

\subsection{Geometric proof}
\begin{figure}
    \centering
    \includegraphics[width=0.6\linewidth]{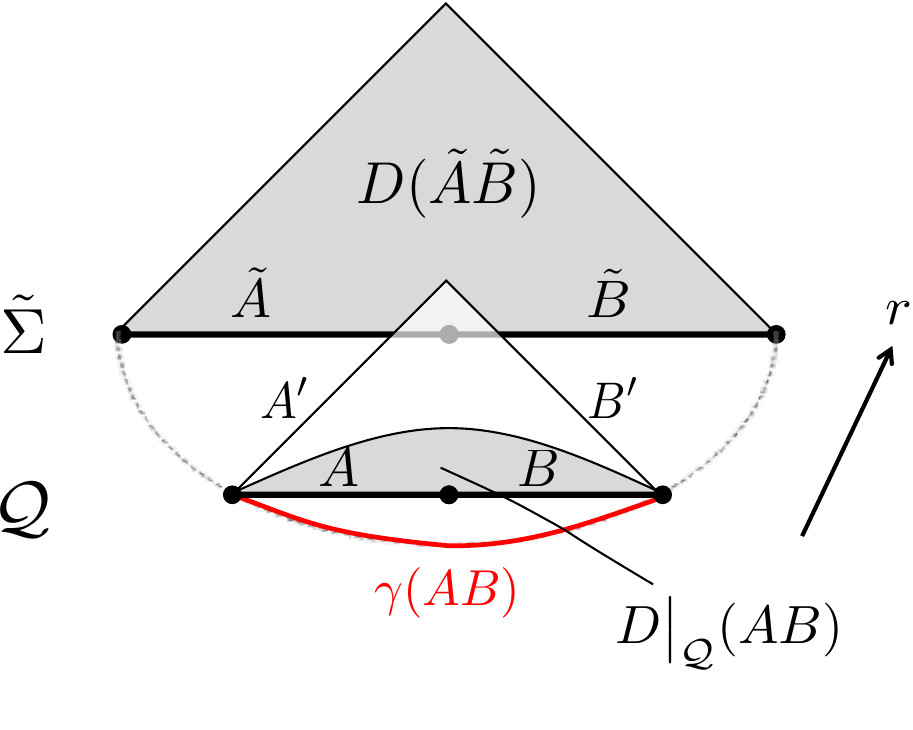}
    \caption{The induced causal diamond and the HRT surface. 
    The induced diamond $D\big|_{\mathcal{Q}} (R)$ is given by an intersection between the causal wedge $\mathcal{C}(\tilde{A}\tilde{B})$ and $\mathcal{Q}$. 
    The HRT surface $\gamma(AB)$ is spacelike separated from $D\big|_{\mathcal{Q}} (AB)$, but is timelike separated from $A'B'$ lying outside of $D\big|_{\mathcal{Q}} (AB)$.}
    \label{fig:SA-geo}
\end{figure}

In this subsection, we present a geometric proof of subadditivity of holographic entanglement entropy for induced causal diamonds. 
The proof follows from a powerful result from~\cite{Grado-White:2020wlb}.
Recall that the standard HRT surface of an interval $R$ can be obtained via the maximin procedure. 
Authors of~\cite{Grado-White:2020wlb} proposed that, when computing holographic entanglement entropy of an interval $R$ on a cutoff surface $\mathcal{Q}$, one should restrict considerations to candidate surfaces that are achronal to $R$.
Indeed, they proved that for \emph{restricted} HRT surfaces, basic quantum information theoretic inequalities, such as subadditivity and strong subadditivity hold. 

With this result in hand, let us revisit the setup from Section~\ref{sec:SA-viol-puzzle} where we demonstrated a violation of subadditivity. 
Note that the HRT surface (or a geodesic $\gamma(AB)$) of an interval $AB$ is on the equal timeslice, and lies very close to $AB$ in the regime with a small tension $T$. 
The boosted interval $A'B'$ is in the future of $AB$, and thus, is timelike separated from the HRT surface $\gamma(AB)$. As such, the restricted HRT surface of $A'B'$ is \emph{not} given by $\gamma(AB)$, which hints that whenever $A'B'$ is not achronal to $\gamma (AB)$, $S_{A'B'}$ may be given by some other curve, and thus $S_{AB}\neq S_{A'B'}$.\footnote{The fact that the interval $A'B'$ is achronal to $\gamma(AB)$ was pointed out in~\cite{Lewkowycz:2019xse} where the authors studied a potential violation of strong subadditivity in the $T\bar{T}$ deformation.}

In the previous subsection, we demonstrated that subadditivity still holds if we boost an interval $AB$ only within the induced causal diamond. 
The connection with the result from~\cite{Grado-White:2020wlb} can be established from the following observation. 
Recall that the induced causal diamond is given by the intersection between the causal wedge $\mathcal{C}(\tilde{A}\tilde{B}) =D(\tilde{A}\tilde{B})$ of an interval $\tilde{A}\tilde{B}$ on the fictitious boundary $\tilde{\Sigma}$ and the cutoff surface combined with the asymptotic boundary $\mathcal{Q}\cup\Sigma$.
In the case of pure AdS, the (static) Ryu-Takayanagi (RT) surface lies exactly on the causal wedge.
Hence, as long as $AB$ is boosted within the induced causal diamond, it is still spacelike separated from the RT surface $\gamma(AB)$, and the result from~\cite{Grado-White:2020wlb} applies. See Fig.~\ref{fig:SA-geo} for the illustration.

This observation can be further extended to show that subadditivity must be obeyed within the induced causal diamond on a generic ground. 
Let us begin by specifying how the induced causal diamond $D\big|_{\mathcal{Q}} (R)$ can be constructed. 
Imagine two points $\tilde{c}_+$ and $\tilde{c}_-$ on an asymptotic boundary $\tilde{\Sigma}$ which are timelike separated. 
Assume that $\tilde{c}_+$ is in the future of $\tilde{c}_-$.
Considering the past/future light cones of $\tilde{c}_+$ and $\tilde{c}_-$, one can define a causal diamond for some region $\tilde{R}$ on $\tilde{\Sigma}$:
\begin{align}
D(\tilde{R}) = J^{-}(\tilde{c}_+) \cap J^{+}(\tilde{c}_-) \cap \tilde{\Sigma}.
\end{align}
An induced causal diamond on $\mathcal{Q}\cup\Sigma$ can be constructed by considering induced light cones from $\tilde{c}_+$ and $\tilde{c}_-$:
\begin{align}
D\big|_{\mathcal{Q}} (R) \equiv J^{-}(\tilde{c}_+) \cap J^{+}(\tilde{c}_-) \cap (\mathcal{Q} \cup \Sigma )
\end{align}
which defines some region $R$ on $\mathcal{Q}\cup\Sigma$ such that 
\begin{equation}
    \partial R = \partial J^{-}(\tilde{c}_+) \cap \partial J^{+}(\tilde{c}_-) \cap (\mathcal{Q} \cup \Sigma ).
\end{equation}
We would like to show that all the intervals within the induced causal diamond $D\big|_{\mathcal{Q}} (R)$ have achronal HRT surfaces. 
This follows from~\cite{Headrick:2014cta,Wall:2012uf,Engelhardt:2014gca} that the entanglement wedge of $R$ must lie outside the causal wedge of $\tilde{R}$, and as such, it is achronal to any points on the induced causal diamond of $R$. 
Hence, we arrive at the following statement. 

\begin{itemize}[leftmargin=*]
\item[] \textbf{Entanglement within induced causal diamond:} Let $D\big|_{\mathcal{Q}} (R)$ be an induced causal diamond on $\mathcal{Q}$ that originates from a causal diamond $D(\tilde{R}) \equiv J^{-}(\tilde{c}_+) \cap J^+(\tilde{c}_-)$, associated with two points $\tilde{c}_+$ and $\tilde{c}_-$ on $\tilde{\Sigma}$.
Then, holographic entanglement entropy within $D\big|_{\mathcal{Q}} (R)$ obeys subadditivity (and strong subadditivity). 
\end{itemize}

In summary, we proved that subadditivity of holographic entanglement entropy remains valid by considering induced light cones.
This provides other evidence supporting our proposal.

In the next section, we turn our attention to yet another puzzle, pointed out by Omiya and Wei~\cite{Omiya:2021olc}, concerning the superluminal signaling in the intermediate picture, and propose a resolution along a similar line of arguments.

\section{On the Omiya-Wei puzzle}\label{sec:OW-puzzle}

\begin{figure}
    \centering    \includegraphics[width=.6\textwidth]{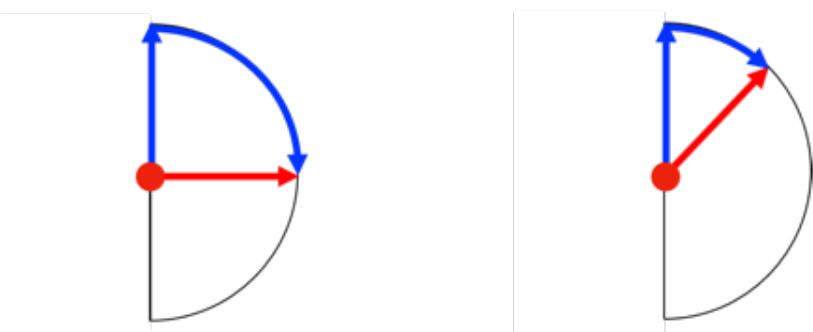}
    \caption{
    The bulk shortcut in the brane-to-boundary signaling.
    The tensionless limit case is shown. This leads to potential superluminal signaling in the intermediate picture.
     The left and right figures show different null trajectories through the bulk. 
     Two excitations {occurred at the same point on the brane} reach $\Sigma$ at the same time in the bulk (illustrated in red) but have different {direction-dependent} superluminal velocities in the intermediate picture (illustrated in blue).}
    \label{fig:brane-signal}
\end{figure}

In this paper, we considered the 2-to-2 scattering in the bulk. 
It turns out that even a ``1-to-1'' scattering, or a brane-to-boundary signaling process, suggests a potential violation of causality, as pointed out in~\cite{Omiya:2021olc}. 
For simplicity, consider a signal emanating from the center of the $T\rightarrow +0$ brane. As shown in Fig.~\ref{fig:brane-signal}, it takes $\pi/2$ time to reach the asymptotic boundary regardless of the direction. On the other hand, the dual trajectory in the intermediate picture would be the blue curves and the time duration is greater than $\pi/2$ assuming the usual causality on the brane. 

One possible resolution of this puzzle, suggested by the authors of~\cite{Omiya:2021olc}, would be that there may exist some nonlocal couplings between the brane $Q$ and the asymptotic boundary $\Sigma$ in the intermediate picture. 
We, however, think that this proposal of nonlocal coupling is not likely the resolution for several reasons.

First, observe that two excitations at the same point on the brane $Q$ appear to have different superluminal velocities, as the superluminal signal propagation speeds on $Q$ would be direction-dependent. 
This observation suggests that nonlocal coupling, if exists, would be a rather fine-tuned and drastic one that connects a point on the brane to every point on the asymptotic boundary. The second argument concerns the holographic quantum task. Suppose that $Q$ and $\Sigma$ are somehow nonlocally coupled, and as a result, one of the inputs, say $c_2$, can be revealed at the location of $c_1$ on $Q$ via the bulk shortcut time.
While this would bring $c_1,c_2$ to the same location, in order to achieve the direct $Q\cup \Sigma$ scattering, one needs to bring them to $r_1,r_2$. 
This, however, does not seem possible.
While we cannot deny the possibility of more intricate use of nonlocal coupling, at least, it does not directly address the origin of the direction dependence.

We propose that this puzzle can be resolved from induced light cones as well since these lead to an apparent superluminal propagation of signals on $Q$. 
Here, the direction dependence of the apparent superluminal velocities is reflected to that in the induced light cones. 
It is worth emphasizing that this apparent superluminal propagation does not imply the violation of the standard causality.
Namely, the apparent nonlocality can arise as a consequence of the state preparation as depicted in Fig.~\ref{fig:loc-vs-nonloc}.

\section{Concluding remarks}\label{sec:proposals}

In this paper, we have discussed three puzzles concerning holography with a brane $Q$ and a cutoff surface $\mathcal{Q}$: i) the disconnected entanglement wedge in the bulk scattering process, ii) a violation of subadditivity of holographic entanglement entropies on a brane, and iii) superluminal signaling between a brane and an asymptotic boundary. 
We then demonstrated that, by considering induced light cones starting behind $Q$ or $\mathcal{Q}$, these puzzles can be avoided. As for i), we presented a generalization of the connected wedge theorem for induced light cones. As for ii), we presented a generic argument that subadditivity remains valid within induced causal diamonds. As for iii), a resolution can be obtained naturally from superluminal light cones. 

One drawback of our proposal is the necessity to (re)introduce the fictitious asymptotically AdS boundary $\tilde{\Sigma}$. Ideally, it would be desirable to construct induced light cones and induced causal diamonds without any reference to $\tilde{\Sigma}$. A potential approach to address this concern is by constructing the induced causal diamond of an interval $R$ in such a way that it remains achronal to the HRT surface of $R$. This approach allows us to solely focus on actual bulk regions without the need to consider regions behind $Q$ or $\mathcal{Q}$.

Here, we conclude this paper with several remarks and speculations.

\subsection{Physical interpretation}

Let us explore potential physical interpretations and implications of our proposal. Recalling the aforementioned puzzles, it is clear that local excitations propagating in the radial direction toward the bulk were causing the trouble. 
On the contrary, local excitations, which propagate on $Q$ (or $\mathcal{Q}$) without leaving it, do not suffer from any causality violation. 
As illustrated in Fig.~\ref{fig:loc-vs-nonloc}, they have rather different profiles of causal diamonds. 
This motivates us to distinguish two types of local excitations in the following manner. 

\begin{itemize}[leftmargin=*]
\item[] \textbf{Trapped excitation:} a local excitation on $Q$ that propagates on $Q$ and does not travel radially to the bulk. This can be created by local operators on $Q$ and obeys the standard causality that is expected from local QFT. In the bulk perspective, this is expected to be a brane-localized field or deficit dual to a special boundary primary operator~\cite{Kanda:2023zse,Miyaji:2022dna,Geng:2021iyq}.

\item[] \textbf{Radially propagating excitation:} a local excitation that travels radially to the bulk. This cannot be created by local operators on $Q$. Namely, creating them requires nonlocal operators, whose complexity is high, and thus, cannot be created instantaneously. In the bulk perspective, this is mediated by the bulk fields, thus a local excitation on $Q$ defined from the extrapolate dictionary~\cite{Porrati:2001gx,Neuenfeld:2021wbl} is of this kind. Accordingly, such excitations have a violation of microcausality between $Q$ and $\Sigma$~\cite{Omiya:2021olc}.
\end{itemize}

Note that a radially propagating excitation has superluminal propagation speed, but this causality violation {on $Q$ or $\mathcal{Q}$ is only an apparent one. 
While this apparent violation of causality is consistent with the HRT formula, it should not appear when we restrict to trapped excitations by local operators on $Q$.
This viewpoint provides an illuminating implication concerning a violation of subadditivity. 
In the current setup with two types of excitations an effective description with the standard causality, which deals only with trapped excitations, may be an incomplete one. 
In order to construct a complete quantum description, one has to consider both types of excitations and thus will need to give up the standard causality.

Based on this observation, we propose that a violation of subadditivity stems from two incompatible assumptions. 
\begin{itemize}[leftmargin=*]
\item[] \textbf{Standard causality on $Q$:} The use of the standard causal diamond would be fine if we restrict our attention to trapped excitations. If we consider both types of excitations as the building blocks of our theory, then we need to give up the standard causality and rely on the induced causal diamond with apparent superluminal signaling.\footnote{A similar line of argument has been made from the perspective of the algebra of causal and entanglement wedges in cutoff holography~\cite{Lewkowycz:2019xse}.} 
\item[] 
\textbf{HRT formula:} The HRT formula computes holographic entanglement entropy for a complete QFT on $Q$ or $\mathcal{Q}$, and as such, considers both types of excitations. As such, its calculation is consistent only with the induced causal diamond.
\end{itemize}

This interpretation seems to resolve all the puzzles mentioned in this paper on an equal footing. 
It will be interesting to test this proposal by direct gravitational or field theoretic calculations.

\subsection{Tensor network interpretation}

\begin{figure}
    \centering
    \includegraphics[width=0.9\textwidth]{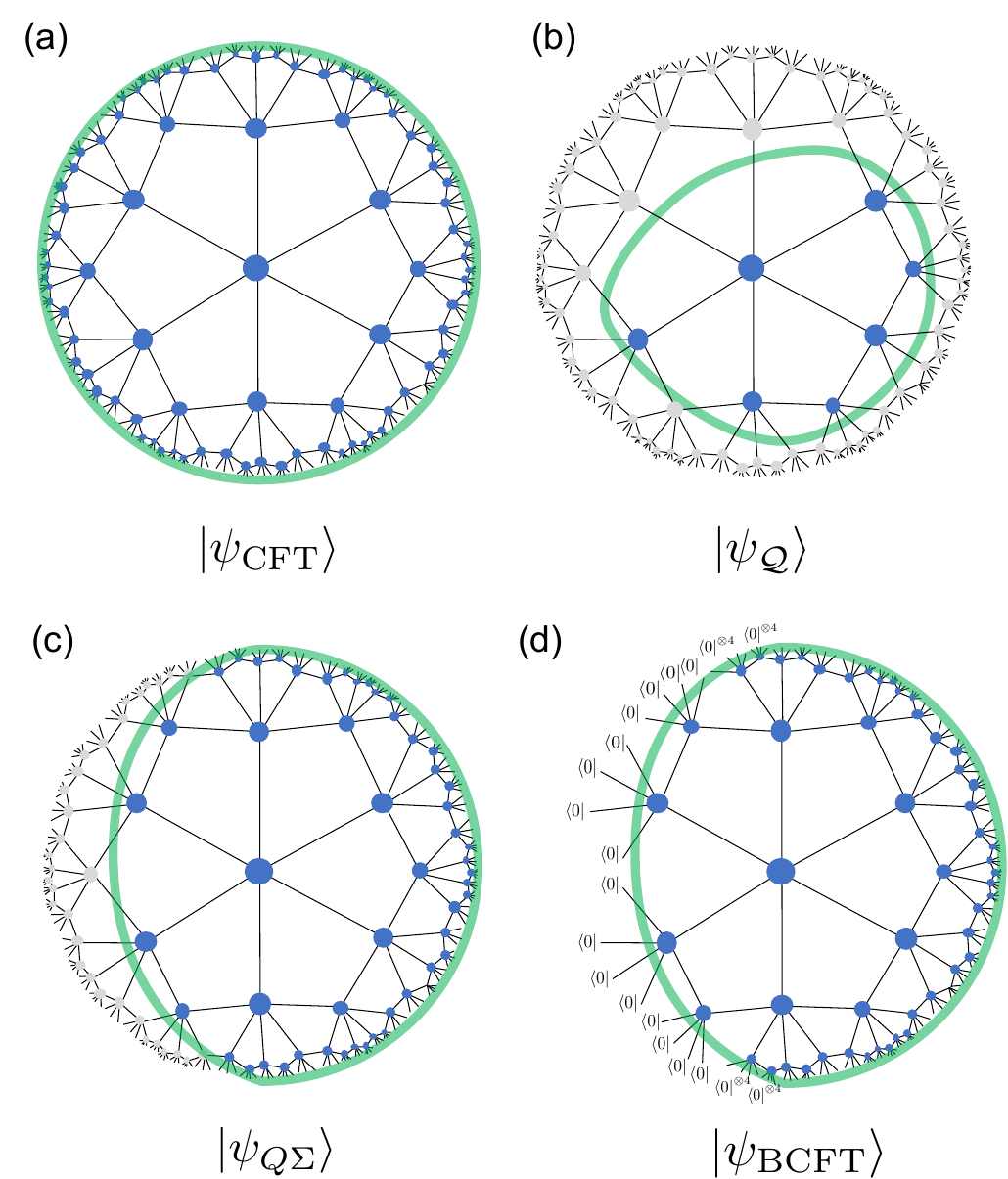}
    \caption{Various holographic tensor network states. Blue dots denote Haar random tensors. These states $\ket{\psi_{\mathrm{CFT}}},\ket{\psi_\mathcal{Q}},\ket{\psi_{Q\Sigma}},\ket{\psi_{\mathrm{BCFT}}}$ have the same RT formula respectively as (a) the CFT ground state, (b) the ground state on $\mathcal{Q}$ dual to a cutoff AdS, (c) the ground state on $Q\cup\Sigma$ in the intermediate picture, and (d) the BCFT ground state. 
    }
    \label{fig:TN-hol}
\end{figure}

Tensor network toy models can provide useful insights into the entanglement structure in holography with $Q$ or $\mathcal{Q}$.
Let us start by briefly reviewing a tensor network model of the AdS/CFT along the lines of~\cite{Hayden:2016cfa} by using Haar random tensors on a hyperbolic lattice (Fig.~\ref{fig:TN-hol} (a)). 
The network defines a wavefunction $|\psi_{\text{CFT}}\rangle$ on the asymptotic boundary and obeys the RT formula at the leading order in $S_{\text{bond}}$, where the bond dimension is given by $e^{S_{\text{bond}}}$. 
The network can be interpreted as a quantum circuit that evolves from the bulk center to the asymptotic boundary. 

Let us consider a setup with the cutoff surface, as well as the intermediate picture in the AdS/BCFT correspondence. As shown in Fig.~\ref{fig:TN-hol} (b), a wavefunction $|\psi_{\mathcal{Q}}\rangle$ on a cutoff surface $\mathcal{Q}$ can be constructed by terminating the tensor network on $\mathcal{Q}$ and leaving tensor legs open.
This procedure is often called the surface/state correspondence~\cite{Miyaji:2015yva}. Due to the use of Haar random tensors, the wavefunction $|\psi_{\mathcal{Q}}\rangle$ on a cutoff surface also obeys the RT formula. 
Namely, holographic entanglement entropy computed for this wavefunction matches with the one that can be obtained from the RT formula applied to the cutoff surface. 
Similarly, by opening the legs on the brane $Q$, one may create a tensor network state $\ket{\psi_{Q\Sigma}}$ corresponding to the intermediate picture (Fig.~\ref{fig:TN-hol}~(c)).
It should be, however, noted that the tensor network model is not capable of accommodating dynamical effects. 
The Neumann boundary condition is sensitive to the gravitational backreaction, as we will discuss in the next section. 

A toy model of holographic BCFT can be constructed by applying a projection operator $\Pi_{Q}=|\phi_Q\rangle\langle\phi_Q|$ on a brane $Q$ and normalizing the outcome:
\begin{equation}
    |\psi_{\mathrm{BCFT}}\rangle \propto \langle\phi_Q| \psi_{Q\Sigma}\rangle.
\end{equation}
In the BCFT, a spatial boundary can be obtained by exchanging $t$ and $x$ of a boundary state which is known to have vanishing entanglement. 
Hence, it is natural to choose $|\phi_{Q}\rangle$ to be a short-range entangled state, such as a product state $|\phi_Q\rangle = |0\rangle^{\otimes n_Q}$ where $n_Q$ represents the number of qubits (or tensor legs) on $Q$ (Fig.~\ref{fig:TN-hol} (d)). 
With this choice of the projection operator $\Pi_Q$, one can verify that the resulting wavefunction $|\psi_{\mathrm{BCFT}}\rangle$ obeys the RT formula of BCFT, where the RT surface terminates on a brane $Q$.

The choice of $|\phi_{Q}\rangle$ being a product state can be also justified by examining whether the RT formula holds or not when $|\phi_{Q}\rangle$ is an entangled state.
Given the RT surface $\gamma(A)$ of $A$ which ends on a brane, one can split a brane $Q$ into two parts $Q_A$ and $Q_{B}$.
In a random tensor network, the entanglement entropy of a subregion is given by the geodesic length plus contributions from the bulk entanglement across the RT surface.
Hence, if $|\phi_Q\rangle$ is entangled across $Q_A$ and $Q_{B}$, then $S(A)$ will be given by the sum of the geodesic length and the entanglement entropy $S_{Q_{A}}$ of $|\phi_Q\rangle$.
Any entanglement in $|\phi_Q\rangle$ leads to a deviation from the RT formula, and as such, one is directed to choose $|\phi_Q\rangle$ to be a product state.

One can also directly relate the BCFT wavefunction $|\psi_{\text{BCFT}}\rangle$ to the CFT wavefunction $|\psi_{\text{CFT}}\rangle$ by using the isometric nature of a random tensor network. 
Let us denote the Hilbert space for $|\psi_{\text{CFT}}\rangle$ as $\mathcal{H}_{\text{CFT}}=\mathcal{H}_{L}\otimes \mathcal{H}_{R}$ by splitting the whole Hilbert space into the left and right parts. 
Recalling that a tensor network realizes an isometry from the bulk to the boundary, a projection operator $\Pi= |\phi_{Q}\rangle\langle \phi_{Q}|$ on $Q$ can be expressed as some projection operator on $\mathcal{H}_{\text{CFT}}$. Since the brane profile $Q$ lies inside the entanglement wedge of $L$, one can write down the projector exclusively on $\mathcal{H}_{L}$. 
Hence, there exists a projector $\Pi_{L}= |\widetilde{\phi_{Q}}\rangle\langle \widetilde{\phi_{Q}}|$ such that 
\begin{align}
(\Pi_{L} \otimes I_{R})|\psi_{\text{CFT}}\rangle_{LR} \propto |\widetilde{\phi_{Q}}\rangle_{L} \otimes |\psi_{\text{BCFT}}\rangle_{R}
\end{align}
which generates $|\psi_{\text{BCFT}}\rangle$ on $\mathcal{H}_{R}$.
This picture is consistent with the conventional interpretation that a brane insertion can be understood as a projection on the CFT wavefunction.

In fact, tensor network toy models provide a useful insight into why one should \emph{not} consider a brane with a negative tension. 
When $T<0$, the brane $Q$ will be located closer to the asymptotic boundary $\Sigma$, and it lies inside the entanglement wedge of $R$. 
This suggests that the projector $\Pi= |\phi_{Q}\rangle\langle \phi_{Q}|$ on $Q$ cannot be represented as a projector acting on $\mathcal{H}_{L}$. 
Instead, in order to generate $|\psi_{\text{BCFT}}\rangle_{R}$
from
$|\psi_{\text{CFT}}\rangle_{LR}$, one has to apply some quantum operation which also acts on $R$ as well. 
Hence, the BCFT wavefunction $|\psi_{\text{BCFT}}\rangle_{R}$ with $T<0$ cannot be generated via a projection acting on the complementary subregion $L$ in the CFT wavefunction $|\psi_{\text{CFT}}\rangle_{LR}$.

Let us conclude by commenting on the problem concerning locality of the Hilbert space on $\mathcal{Q}$. 
Whether the standard locality remains valid on a brane and a cutoff surface has been occasionally questioned in the literature. 
Indeed, one might think that the aforementioned puzzles could also be understood as further hints of a breakdown of locality. 
Tensor network toy models present an intriguing perspective on locality of the Hilbert space on $\mathcal{Q}$. 
Indeed, given a wavefunction $|\psi\rangle$ of the dual CFT, one can naturally introduce a cutoff surface $\mathcal{Q}$ and define an induced wavefunction $|\psi_\mathcal{Q}\rangle$ on $\mathcal{Q}$ without spoiling the local Hilbert space structure.
Also, tensor network calculations suggest that, at least for wavefunctions where the holographic entanglement entropy formula applies, the standard local Hilbert space with a small cutoff should be present as the projected state $|\phi_\mathcal{Q}\rangle$ should carry short-range entanglement only. For these reasons, we attempt to resolve the causality puzzles concerning a brane and a cutoff surface without giving up the standard locality in this paper. 

\subsection{Boundary conditions}\label{sec:bdy-cond}

The crux of our proposal is that local excitations, namely radially propagating ones, cannot be created by local operations. 
Understanding the potential physical mechanisms behind such a restriction requires us to revisit the problem of choices of boundary conditions.
Indeed, two boundary conditions suggest different mechanisms for such restrictions. 

Let us begin with the Neumann boundary condition. 
In this case, nonvanishing graviton fluctuations are allowed which we have completely ignored until this point in this paper.
The graviton effect will become significant when the brane $Q$ is perturbed by matter fields and its location starts to move. 
When the tension $T$ is small, the brane $Q$ will wobble around violently even under a tiny perturbation.
This type of behavior has been studied in~\cite{Chen:2020uac,Hernandez:2020nem}. 
It is associated with the breakdown of the local QFT description on $Q$. 
These previous works naturally lead us to propose that a radially propagating excitation is prohibited because a local perturbation significantly backreacts to the brane $Q$.
Furthermore, we speculate that a local excitation may be introduced without backreaction by careful fine-tuning (\emph{e.g.} by adiabatically inserting an excitation), but this will require long preparation, and as such, the causality violation can be avoided.

Next, let us turn to the Dirichlet boundary condition. 
One useful interpretation is to view the wavefunction $|\psi_{\mathcal{Q}}\rangle$ as a coarse-grained wavefunction of $|\psi_{\mathrm{CFT}}\rangle$ where some irrelevant information in $|\psi_{\mathrm{CFT}}\rangle$ was thrown away by gradually changing the length scale of interest.
In this interpretation, one can naturally associate the cause of apparent nonlocality of a radially propagating excitation to degrees of freedom that have been thrown away.
Indeed, we speculate that some of such information should not have been forgotten in order to describe the physics of these excitations. In our proposal of induced light cones, they are first prepared on the asymptotic boundary and spread behind $\mathcal{Q}$ until their induced light cones reach $\mathcal{Q}$. 
The cutoff surface $\mathcal{Q}$ seems to trace out these initial preparations for creating those radially propagating excitations.
One concrete way to obtain a causally sensible quantum description is to restore the coarse-grained degrees of freedom back. 
They may act as an environment, or a bath, for degrees of freedom on $\mathcal{Q}$, and assist to generate apparent superluminal signaling. 

\acknowledgments

We thank Alex May, Rob Myers, Dominik Neuenfeld, Valentina Prilepina, Shang-Ming Ruan, Manu Srivastava, Tadashi Takayanagi, Yuya Kusuki, and Zixia Wei for their useful discussions and comments. 
T.M.'s stay at Perimeter Institute was supported in part by SOKENDAI Student Dispatch Program. 
T.M. is grateful for the hospitality of Perimeter Institute where part of this work was carried out. Research at Perimeter Institute is supported in part by the Government of Canada through the Department of Innovation, Science and Economic Development and by the Province of Ontario through the Ministry of Colleges and Universities.
T.M. was partially supported by SOKENDAI and the Atsumi Scholarship from the Atsumi International Foundation. This work was supported by JSPS KAKENHI Grant Number 23KJ1154.

\appendix
\section{Calculation of holographic entanglement entropy}\label{app:HEE}
In this section, we present formulae for geodesics calculating holographic entanglement entropy for subregions on a brane $Q$ and the asymptotic boundary $\Sigma$.

Since the background is static, we employ the RT prescription for holographic entanglement entropy.
The geodesic between two spacelike points $X_A$ and $X_B$ in the embedding coordinates is given by
\begin{equation}
    d(A:B) = R\, \mathrm{arccosh}\,\zeta,
    \label{eq:geodesic-gen}
\end{equation}
where\footnote{The inner product is defined as $X\cdot Y = -X_0 Y_0 -X_3 Y_3 + X_1 Y_1 +X_2 Y_2$.}
\begin{equation}
    \zeta=-\frac{X_A\cdot X_B}{R^2}.
    \label{eq:zeta}
\end{equation}
When $\zeta\gg 1$ (\emph{e.g.} when at least one of the two points is on the asymptotic boundary), \eqref{eq:geodesic-gen} reduces to
\begin{equation}
    d(A:B)=R\log\qty(\zeta+\sqrt{\zeta^2-1})\approx R\log(2\zeta).
    \label{eq:geodesic-bdy-lim}
\end{equation}

In the brane coordinates $(t,\eta)$, the embedding coordinates on $Q$ are given by
\begin{equation}
    \begin{split}
    \sqrt{R^2+r^2} &= \frac{X_0}{\cos t} = \frac{X_3}{\sin t}= R \frac{\sqrt{1+\lambda^2}}{\abs{\sin\eta}} \nonumber \\
     X_1 &= -\lambda R,  \nonumber\\
     X_2 &= \lambda R \sqrt{1+\lambda^{-2}} \cot \eta.
    \end{split}
    \label{eq:brane-embed}
\end{equation}

For two points in global coordinates $(t_A,r_A,\theta_A),(t_B,r_B,\theta_B)$, 
\begin{equation}
    \zeta = \sqrt{\left(1+\frac{r_A^2}{R^2}\right)\left(1+\frac{r_B^2}{R^2}\right)} \cos(t_A-t_B) - \frac{r_A r_B}{R^2}\cos(\theta_A-\theta_B).
    \label{eq:globalAdS-geodesic}
\end{equation}

When the two endpoints are on the asymptotic boundary $\Sigma:r=r_\infty\gg R$, \eqref{eq:globalAdS-geodesic} becomes
\begin{equation}
    \zeta_\Sigma = \frac{r_\infty^2}{R^2}\qty[ \cos(t_A-t_B) - \cos(\theta_A-\theta_B)].
    \label{eq:zeta-discon-bdy}
\end{equation}

When the two endpoints are on the ETW brane $Q$, we can apply \eqref{eq:brane-embed} to \eqref{eq:zeta} to obtain $\zeta$ in the brane coordinates $(t_A,\eta_A)$ and $(t_B,\eta_B)$:
\begin{equation}
    \zeta_Q=(1+\lambda^2)\frac{\cos(t_A-t_B)-\cos(\eta_A-\eta_b)}{\sin\eta_A\sin\eta_B}+1.
    \label{eq:zeta-discon-Q}
\end{equation}
We removed the absolute value symbols of $\sin\eta_A$ and $\sin\eta_B$ as they have the same sign whenever there is only a single brane.

When one endpoint $(t_\Sigma,r_\infty,\theta)$ lies on $\Sigma$ and the other $(t_Q,\eta)$ lies on $Q$, \eqref{eq:brane-embed} becomes
\begin{align}
    \zeta_{\Sigma Q}
    &= \frac{r_\infty}{R}\qty[\sqrt{1+\lambda^2}\frac{\cos(t_\Sigma-t_Q)-\cos\theta\cos\eta}{\abs{\sin\eta}} +\lambda \sin\theta].
    \label{eq:zeta-con}
\end{align}
We used that the sign of $\lambda$ and $\sin\eta$ is by definition opposite.

\section{The UV cutoffs on the brane in the boundary limit}\label{app:cutoff}
In this section, we consider approaching the asymptotic boundary in various ways and relating each coordinate to the UV cutoff on $\Sigma$.

The brane coordinate $\eta$ approaches the asymptotic boundary $\Sigma$ in the limit ${\eta\rightarrow 0,\pi}$. Denoting this asymptotic value of $\eta$ by $\eta_\infty$ and $\theta$ by $\theta_\infty$, they are related to $R^2/r_\infty$, the UV cutoff on $\Sigma$, as
\begin{equation}
    \frac{1+\lambda^2}{\sin^2\eta_\infty}=1+\frac{r_\infty^2}{R^2}=\frac{\cos^2\theta_\infty}{\cos^2\eta_\infty},\quad -\lambda = \frac{r_\infty}{R}\sin\theta_\infty.
\end{equation}

We can also consider the asymptotic boundary as the brane with a critical tension $T\rightarrow 1\Leftrightarrow \lambda\rightarrow\infty$. By denoting $\lambda$ in the critical limit by $\lambda_\infty$, it is related to $r_\infty$ as
\begin{equation}
    \lambda_\infty=-\frac{r_\infty}{R}\sin\theta.
\end{equation}
Note that $\eta=\theta$ in the critical limit.

\bibliographystyle{JHEP}
\bibliography{qtask}

\end{document}